\documentclass[aps,prx,twocolumn,groupedaddress,showpacs,preprintnumbers,amsmath]{revtex4-1}
\usepackage{epsfig,amsmath,amssymb,graphics,color,calc,hyperref}
\usepackage{mathtools}
\newcommand\underrel[2]{\mathrel{\mathop{#2}\limits_{#1}}}

\begin{document}

\title{The Many-Body localization transition in the Hilbert space}

\author{M. Tarzia}

\affiliation{LPTMC, CNRS-UMR 7600, Sorbonne Universit\'e, 4 Pl. Jussieu, F-75005 Paris, France\\
Institut  Universitaire  de  France,  1  rue  Descartes,  75231  Paris  Cedex  05,  France}

\begin{abstract}
In this paper we propose a new perspective to analyze the many-body localization (MBL) transition when recast in terms of a single-particle tight-binding model in the space of many-body configurations.
We compute the distribution of tunneling rates between many-body states separated by an extensive number of spin flips 
at the leading order in perturbation theory starting from the insulator, and determine the scaling of their typical amplitude with the number of accessible states 
in the Hilbert space. 
By using an analogy with the Rosenzweig-Porter random matrix ensemble, we propose an ergodicity breaking criterion for the MBL transition based on the Fermi Golden Rule. According to this criterion, in the MBL phase 
many resonances are formed at large distance 
from an 
infinite temperature initial state, but they are not enough for the quantum dynamics to decorrelate from it in a finite time.
This implies that, differently from Anderson localized states, in the insulating phase many-body eigenstates are multifractal in the Hilbert space, as they occupy a large but subexponential part of the total volume, in agreement with recent numerical results, perturbative calculations, and intuitive arguments. Possible limitations and implications of our interpretation are discussed in the conclusions.
\end{abstract}

\pacs{}

\maketitle



\section{Introduction}

Quantum systems of interacting particles subject to sufficiently strong disorder will fail to come to thermal equilibrium when they are not coupled to an external bath even though prepared with extensive amounts of energy above their ground states. 
This  phenomenon, commonly referred to as Many-Body Localization (MBL),
was originally predicted by Anderson~\cite{anderson}, but firmly established only during the last 15 years, after the famous breakthrough of ~\cite{BAA,Gornyi}, 
and corresponds to a novel dynamical out-of-equilibrium quantum phase transition due to the interplay of disorder, interactions, and quantum fluctuations~\cite{reviewMBL,reviewMBL2,reviewMBL3,reviewMBL4,reviewMBL5}. 
Its existence has received support from perturbative~\cite{BAA,Gornyi}, numerical~\cite{Huse,pal,alet}, and experimental studies~\cite{experiments1,experiments2,experiments3,experiments4,experiments5,experiments6}, as well as rigorous mathematical approaches~\cite{LIOMS}.
These investigations have shown that the main feature of the MBL phase is a robust effective integrability~\cite{LIOMS,LIOMSb,LIOMS1,LIOMS2}: an extensive set of quasilocal integrals of motion emerges, providing an intuitive explanation of the breakdown of thermalization, 
and producing several unusual and remarkable consequences, such as the absence of dc transport~\cite{BAA}, the violation of the eigenstate thermalization hypothesis~\cite{ETH} along with common concepts of equilibrium statistical mechanics~\cite{violation}, and the area-law entanglement of eigenstates~\cite{entanglement,entanglement1,LIOMS1}.

In the latest years these remarkable phenomena have attracted a huge interest (see~\cite{reviewMBL,reviewMBL2,reviewMBL3,reviewMBL4,reviewMBL5} for recent reviews), predominantly from the fact that MBL can protect quantum correlations from decoherence even at finite energy density and for arbitrarily long times. Yet, despite an impressively wide amount of work and several significant progress, many important problems remain open, especially concerning the critical properties of the transition~\cite{critical,thiery,KT,KT1}, the existence of MBL in higher dimensions $d>1$~\cite{avalanches,gopala,doggen}, and the anomalous diffusion and out-of-equilibrium relaxation observed in the ``bad metal'' regime preceding the MBL~\cite{bad_metal1,bad_metal2}.
In this context, simplified effective models might naturally play an important role to sharpen these questions and provide a playground to explore the nature of MBL and improve our understanding of it. 

In this respect, a paradigmatic route which gives a very intuitive picture of MBL is obtained by recasting the many-body quantum dynamics in terms of a single-particle tight-binding problem in the Hilbert space (HS)~\cite{dot,spectral_diffusion}. Within this mapping many-body configurations are seen as site orbitals on a given graph (with strongly correlated diagonal disorder) and the interactions play the role of an effective hopping connecting them. 
Although the local structure and topology of the graph depend on the specific form of the many-body Hamiltonian and on the choice of the basis, the HS is generically a very high dimensional disorder lattice. 
 It is therefore tempting to argue that single-particle Anderson localization (AL) on the Bethe lattice~\cite{abou} can be thought as a pictorial representation of MBL, as put forward in the seminal work of~\cite{dot}, and later further investigated in Refs.~\cite{Jacquod,scardicchio_bethe_mbl,dinamica,logan1}. On the one hand, several intriguing observations support this analogy: The critical point of the MBL transition is expected to be in the localized phase, as for the Anderson model in $d \to \infty$~\cite{SUSY_Bethe,large_d}; Recent phenomenological renormalization group (RG) approaches~\cite{KT,KT1} predict KT-like flows for the MBL transition characterized by two relevant localization lengths (the typical and the average one) which are expected to exhibit the same critical behavior as in AL on the Bethe lattice~\cite{lemarie}; Moreover,  finite-size effects close to the Anderson transition on the Bethe lattice reveal a non-monotonicity~\cite{mirlin} which is also characteristic of 
 the MBL transition.
 
 On the other hand, however, 
there is also a major 
difference regarding the spatial extension of many-body wavefunctions 
in the insulating regime. 
 The statistics of eigenstates is generally characterized by their fractal dimensions $D_q$, defined through the asymptotic scaling behavior of the moments 
 $\Upsilon_q = \langle \sum_i | \psi_\alpha (i) |^{2 q} \rangle$ with the size of the accessible volume ${\cal V}$:  $\Upsilon_q \simeq {\cal V}^{D_q(1 - q)}$ [$q=2$  recovers  the  usual  inverse  participation (IPR)].
For a perfectly delocalized ergodic states similar to plane waves $D_q = 1$.  Conversely, if a state is localized on a finite volume, one gets $D_q = 0$, as for single-particle AL. In an intermediate  situation,  wavefunctions  are  extended  but nonergodic, with $0<D_q<1$.
In contrasts  with  the  well established  case  of  AL,  where  the  spatial  extension  of single-particle orbitals is known to display genuine multifractality {\it only} at criticality (in any dimension $d\ge2$~\cite{RRGmulti}), recent numerical results~\cite{mace,alet,laflorencie}, as well as perturbative calculations~\cite{resonances}, strongly indicate that the many-body wavefunctions are multifractal in the whole MBL regime. 
This result can be easily rationalized considering for instance a quantum spin chain at strong disorder: Deep in the insulating regime most of the spins are strongly polarized 
due to the random fields (paraller or antiparallel to it depending whether one is looking at the ground state or excited states); Yet a small but finite fraction $\rho$ 
of them 
remains ``active'' on the sites where the random field is smaller than the typical tunneling rate 
for spin flips~\cite{mace,resonances,laflorencie}. 
For a chain of length $n$, many-body wavefunctions then typically occupy $2^{n \rho}$ configurations in the spin configuration basis, which yields $D_2 \approx \rho$. 
	Note that at strong disorder the fraction of active spins is proportional to $\Gamma / h$~\cite{alet,resonances}, where $\Gamma$ is the transition rate for spin flips and $h$ is the width of the disorder distribution (see also Sec.~\ref{sec:SD}), and thus is expected to vanish only in the limit of infinite disorder.

Another way of rephrasing the same concept is as follows: In absence of interactions, single-particles orbitals are Anderson localized over a localization length $\xi_{\rm loc}$. Once the interactions are turned on, 
 at strong enough disorder many-body eigenstates are expected to be weak modifications of the Slater determinant of single-particles orbitals. 
 In other words, they must be eigenstates of  the LIOMS~\cite{LIOMS,LIOMSb,LIOMS1,LIOMS2}, which 
 are essentially linear combinations of  local operators 
 over a finite length of order $\xi_{\rm loc}$. Hence, 
 in the Fock space  of the occupation number of single-particle localized orbitals  they roughly occupy a volume of the order order $\xi_{\rm loc}^n$, yielding roughly $D_2 \approx \log_2 \xi_{\rm loc}$~\cite{entanglement_D2}.

These simple arguments indicate that many-body wavefunctions cannot be Anderson localized on a finite volume (except at infinite disorder) and are thus generically nonergodic multifractal in configuration space (with fractal dimensions $0< D_q  < 1$), although for a rather simple reason.
Of course, the localization properties of many-body eigenstates depends crucially on the choice of the basis. However, this is true also for single-particle Anderson localization (e.g., fully delocalized eigenvectors in real space are fully localized in momentum space).  
	The results of Ref.~\cite{mace} clerly indicate  that the multifractal nature of many-body eigenstates  is the same for two 
	relevant choices of the basis that diagonalize the many-body Hamiltonian in specific limits in which the system is completely localized, and are thus used as a starting point for the $l$-bits construction or efficient numerical simulations of 
	MBL.

In this paper we put forward a novel perspective to analyze the 
MBL transition in the HS, providing a clear explanation of the difference between AL and MBL. We focus on 
the non-local propagator 
${\cal G}_{0,nx}$ which connect a $T=\infty$ randomly chosen initial state $| 0 \rangle$ (e.g., in the middle of the many-body spectrum) with the configurations at large extensive distance $nx$ from it (separated by $nx$ spin flips, with $0<x<1$). These matrix elements encode the probability that a system being in the state $| 0 \rangle$  at $t=0$ is found in a state which differs from it by $nx$ spin flips after infinite time.
We evaluate these amplitudes using the Forward Scattering Approximation (FSA)~\cite{anderson,LIOMSb,LIOMS2,pietracaprina}, i.e., at the leading order in perturbation theory starting from the insulator, and analyze the asymptotic scaling behavior of their typical value  with the number ${\cal N}_{nx}$ of accessible configurations at distance $nx$ from $| 0 \rangle$ and in the same energy shell, obtaining $| {\cal G}_{0,nx}^{\rm FSA} |_{\rm typ} \propto({\cal N}_{nx})^{-\gamma/2}$. The exponent $\gamma$  increases with the disorder strength. Based on an analogy with the so-called Rosenzweig-Porter (RP) random matrix ensemble~\cite{kravtsov}, and following the ideas of~\cite{bogomolny,nosov,kravtsov1}, we  put forward  a criterion for ergodicity breaking based on the Fermi Golden Rule (FGR): 
If $\gamma < 1$ the escape rate of the initial state $| 0 \rangle$ 
is much larger than the spread of energy level due to disorder and the system is in the fully ergodic phase; Conversely, if $\gamma >1$ the initial state only hybridize with a sub-extensive fraction  
${\cal V}^{1 - \gamma}$ of the total configurations, 
thereby producing nonergodic multifractal wavefunctions which only occupy ${\cal V}^{2 - \gamma}$ configurations close in energy~\cite{caveat}.
According to this interpretation, although in the MBL phase {\it many} resonances are formed at large distance in the configuration space, they are not enough to ensure ergodicity and to allow  the quantum dynamics to decorrelate from the initial condition in a finite time.
From the point of view of single particle hopping in the HS, MBL is thus reminiscent of the transition from ergodic to multifractal states of the  RP random matrix ensemble~\cite{kravtsov} (and its generalizations~\cite{nosov,kravtsov1}), 
and {\it not} to the transition to AL, which instead occurs at $\gamma = 2$, corresponding to the requirement that the number of resonances found from $| 0 \rangle$ stays {\it finite} in the thermodynamic limit~\cite{bogomolny,nosov,kravtsov1}.
In fact we find that $\gamma$ tends to $2$ from below in the limit of infinite disorder (which can be treated analytically within the FSA), implying that the many-body wavefunctions become truly Anderson localized on a finite volume of the HS only when the density of ``active spins'' vanishes and the bare localization length $\xi_{\rm loc} \lesssim 1$ (i.e., when the LIOMS $\tau_i^z \simeq S_i^z$), in agreement with the intuitive arguments given above.

We apply this approach to three different $1d$ models commonly used in the context of MBL~\cite{pal,alet,znidaric,mace,laflorencie,serbyn,doggen_sub,LIOMS,abanin,huseQP,roscilde,barlev}, showing that 
the ergodicity breaking criterion $\gamma_c=1$ yields an estimation of the critical disorder $h_c$ in strikingly good agreement with the one obtained from the most recent numerical studies of systems of approximately the same size of the ones considered here. This observation supports the robustness of our conclusions and the validity of the ergodicity breaking criterion based on the FGR.
We also backup the perturbative analysis by inspecting the signature of nonergodic multifractal eigenstates by probing the non standard scaling limit of the spectral statistics in exact diagonalizations of the many-body Hamiltonians of small sizes~\cite{facoetti}. 

All in all, our interpretation fully supports the picture recently proposed in Ref.~\cite{laflorencie} where MBL is seen as a fragmentation of the HS (see also Refs.~\cite{chalker} where similar ideas were promoted to explore the analogy between MBL  and a percolation transition in the configuration space).

The paper is organized as follows. In the next section we define the model and its HS representation as a single-particle tight-binding problem. In Sec.~\ref{sec:FSA} we present the results obtained within the FSA for the scaling of the matrix elements between distant states in the HS. In Sec.~\ref{sec:RP} we examine the analogy with the RP model and discuss the ergodicity breaking criterion based on the FGR in the light of this analogy. In Sec.~\ref{sec:Bethe} we recall the results obtained for  the Anderson model on the Bethe lattice within the FSA and analyze the differences between AL and MBL. In Sec.~\ref{sec:SD} we present a strong disorder approximation 
for $\gamma$. In Sec.~\ref{sec:ldos} we study the signature of the presence of multifractal states in the anomalous scaling limit of the local spectral statistics obtained from exact diagonalizations of small systems. Finally, in Sec.~\ref{sec:2d} we discuss  
the limitations of our interpretation and  in Sec.~\ref{sec:conclusions} we describe some possible implications and perspectives for future investigations.
In App.~\ref{app} we provide more details and supplemental information related to several points discussed in the main text.


\section{The models and Hilbert space representation} \label{sec:model}
We perform our analysis for three paradigmatic $1d$ models for MBL, 
namely the random-field Heisenberg XXZ spin chain, which has been used as a prototype for the MBL transition~\cite{pal,alet,znidaric,mace,laflorencie,serbyn,doggen_sub,chalker,devakul}, the  ``Imbrie'' model, for which the existence of the MBL transition has been proven rigorously~\cite{LIOMS}, 
and a model of interacting (spinless) fermions in a quasiperiodic (QP) potential~\cite{huseQP,roscilde,barlev}, similar to the one actually realized in cold atom experiments~\cite{experiments1,experiments2,experiments3}. In the main text we will mostly focus on the disorder XXZ spin chain, although our results are valid for all the tree models, as shown in App.~\ref{app}.
The Hamiltonian of the random-field XXZ $S=1/2$ chain is:
\begin{equation} \label{eq:Hxxz}
{\cal H}_{\rm XXZ} = \sum_{i=1}^n \left( S_i^x S_{i+1}^{x} + S_i^y S_{i+1}^{y} +\Delta S_i^z S_{i+1}^{z} + h_i S_i^z \right) \, ,
\end{equation} 
with periodic boundary conditions and $h_i$ independent and identically distributed uniformly in the interval $[-h,h]$.
This model has been intensively studied~\cite{pal,alet,znidaric,mace,laflorencie,serbyn,doggen_sub,chalker,devakul} and its phase diagram is known for $\Delta=1$, where a MBL transition takes place at a critical disorder within the interval $h_c \in [3.7,4.5]$ in the middle of the many-body spectrum, $E_0 \approx - n /(4 n - 4)$, in the zero magnetization sector $\sum_{i=1}^n S_i^z = 0$, and for $n \lesssim 24$~\cite{alet,devakul}.

\begin{figure}
\includegraphics[width=0.495\textwidth]{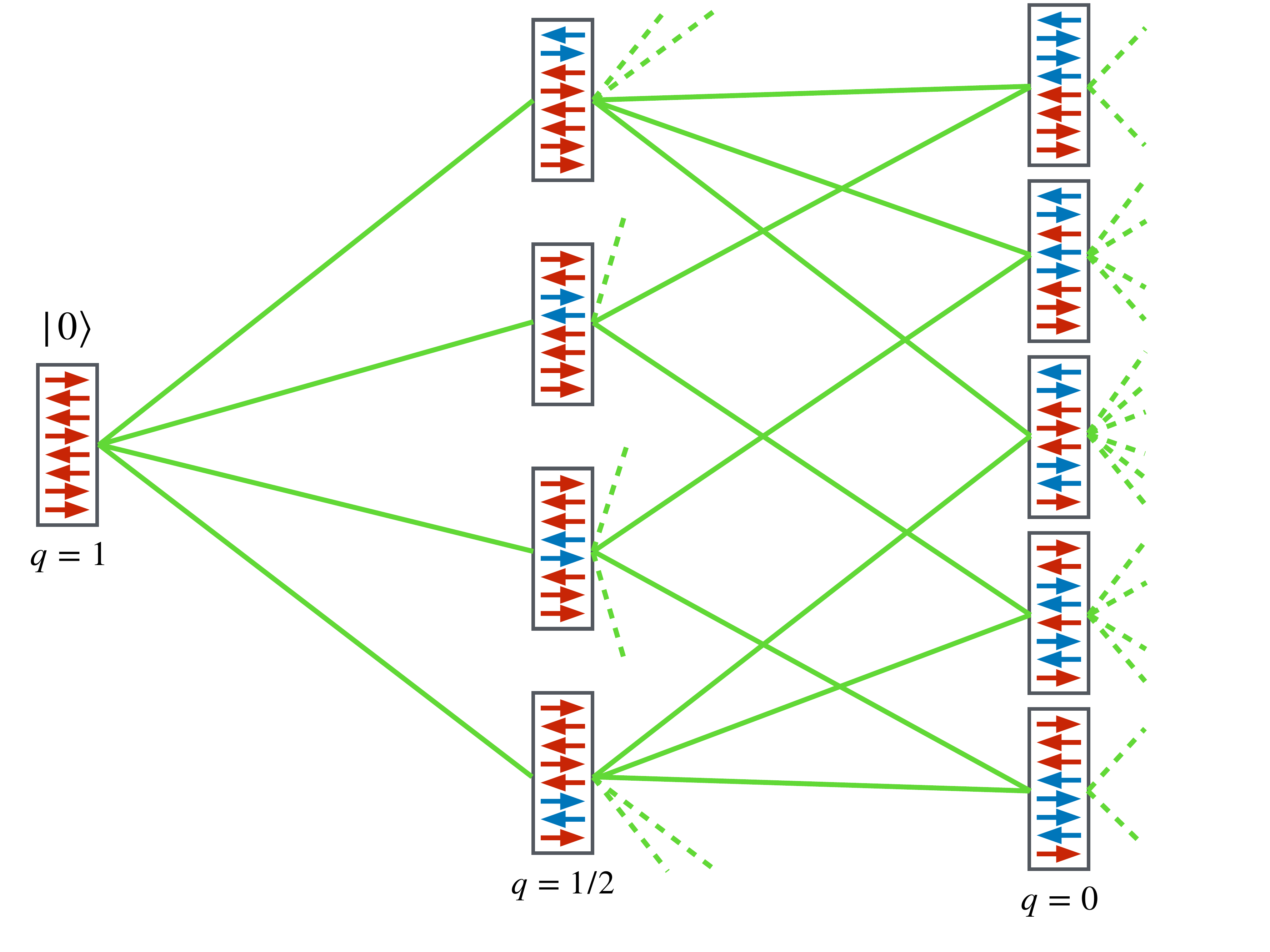}
\caption{\label{fig:graph}
Sketch of the portion of the HS in the spin configuration basis within Hamming distance $n/4$ (i.e., overlap $q\ge 0$) from the state $|0 \rangle$, accessible within the FSA (retaining only the lowest order terms in the locator expansion) for the XXZ random-field spin chain~(\ref{eq:Hxxz}). The blue spins are the ones that have been flipped with respect to the initial configuration. The green lines represent links between many-body configurations connected by the interacting term of~(\ref{eq:Hxxz}), with effective hopping rate $t=1/2$. The green dashed lines represent links to configurations belonging to paths on which spins are flipped more than once and which contribute at higher order in perturbation theory, $O(t^{n/4+2})$ or higher.}
\end{figure}

By choosing as a basis of the HS the simultaneous eigenstates of the operators $S_i^z$, the Hamiltonian~(\ref{eq:Hxxz}) can be recast as a single-particle Anderson problem of the form
\begin{equation} \label{eq:H1}
{\cal H} = \sum_{a=1}^{\cal V} E_a | a \rangle \langle a | + t \sum_{\langle a, b \rangle} \left( | a \rangle \langle b | + | b \rangle \langle a |  \right) \, ,
\end{equation} 
where site orbitals represents many-body states in the spin configurations basis, $| a \rangle = | \! \! \uparrow \downarrow \uparrow \! \cdots \rangle$, the constant hopping rate $t=1/2$ allows tunneling between states
$a$ and $b$ connected by the last term of~(\ref{eq:Hxxz}) which produces spin flips of two neighboring spins of opposite sign. The sums run over all ${\cal V} = {{n}\choose{n/2}}$ many-body configurations with zero magnetization. 
 The diagonal part of~(\ref{eq:Hxxz})  yields the on-site random energies $E_a = \langle a | \sum_{i=1}^n ( \Delta S_i^z S_i^{z+1} + h_i S_i^z)  | a \rangle $, which are strongly correlated~\cite{logan} (the ${\cal V}$ random energies are linear combination of $n$ iid random numbers only).
Note that the spin configuration basis diagonalize ${\cal H}_{\rm XXZ}$ in the infinite disorder limit $h \to \infty$, where all many-body eigenfunctions are completely localized on single sites $| a \rangle$, and is thus suitable to study the stability of the insulating phase. The connectivity of the state $| a \rangle$ is equal to the number of domain walls 
($\uparrow \downarrow$ or $\downarrow \uparrow$) in the spin configuration, and ranges from $2$ (for the $n$ configurations with $n/2$ consecutive $+1/2$ spins and $n/2$ consecutive $-1/2$ spins) to $n$ (for the two Neel states), with average value $\langle z \rangle \approx (n+1)/2$. Hence the network is sparse and high-dimensional, however, differently from the sparse random lattices, it is a deterministic graph with many regular local motifs and loops of all sizes (see Fig.~\ref{fig:graph}).

\section{The Forward Scattering Approximation} \label{sec:FSA}
A simple and powerful route to study and understand single-particle AL is by studying the convergence of perturbation theory starting from the insulator via the so-called the locator expansion~\cite{anderson}.
As shown by Anderson, in the insulating phase of the Anderson model on a $d$-dimensional lattice resonances do not proliferate at large distance in space and the locator expansion converges, implying that hopping only hybridize degrees of freedom within a finite volume of size $\xi_{\rm loc}$.
The FSA  consists in retaining only the lowest order terms in the locator expansion, which amounts in summing only over the amplitudes of the shortest paths connecting two points.
Within this approximation the non-local 
propagator (at energy $E$) between two points $a$ and $b$ at a given distance 
reads:
\begin{equation} \label{eq:FSA}
{\cal G}_{a,b}^{\rm FSA} (E) = \sum_{\textsf{p} \in \textsf{FSP}_{(a,b)}} \prod_{c \in \textsf{p}} \frac{t}{E - E_c} \, ,
	\end{equation}
	where $\textsf{p} \in \textsf{FSP}_{(a,b)}$ denotes a path among all Forward Scattering Paths connecting $a$ and $b$.
Since at strong disorder the amplitude of each path decreases exponentially with its length, this approximation is expected to become more and more accurate as the disorder is increased.

In this paper we use the FSA to estimate the propagators ${\cal G}_{0,nx}$ for the XXZ disordered chain~(\ref{eq:Hxxz}) when recast as a single-particle tight binding problem~(\ref{eq:H1}) in the spin configuration basis. We pick an infinite temperature many-body state $| 0 \rangle = \{ {}^{0}\!S_i^z 
\}$ 
[with energy $E_0 \approx -n/(4n-4)$] and determine the probability distribution of the matrix elements with configurations at distance $nx$ from it (which differ by $nx$ spins) at the lowest order in the ``hopping'' (i.e., the off-diagonal part of the many-body Hamiltonian in the spin configuration basis).

Of course one might argue that the FSA is a crude approximation for the true propagator. 
	However, on the one hand the FSA (i.e., calculating the Green's function by retaining only the lowest order in the off-diagonal terms) has been already successfully applied several times in the context of MBL,  yielding reasonably accurate  estimations of the boundaries of the insulating phase~\cite{baldwin,pietracaprina}, and providing an approximate strategy to construct the LIOMS~\cite{LIOMSb,LIOMS2}.
	Furthermore, since the rigorous results of~\cite{LIOMS} ensure that the perturbative expansion is in fact convergent in the MBL regime, taking only the leading order terms should not provide a too unreasonable starting point at least at strong enough disorder, as demonstrated by the recent quantitative analysis of Ref.~\cite{colmenarez}. However, since one of the main problem of the FSA is that the non-renormalized perturbative expansion has poles for any value of the energy within the support of the probability distribution of the random energies, in the following we will only focus on the {\it typical} value of the propagator, neglecting the effect of large matrix elements in the tails of the distribution. The limitations of this approach will be discussed further in Sec.~\ref{sec:2d}.
	
\begin{figure}
\includegraphics[width=0.468\textwidth]{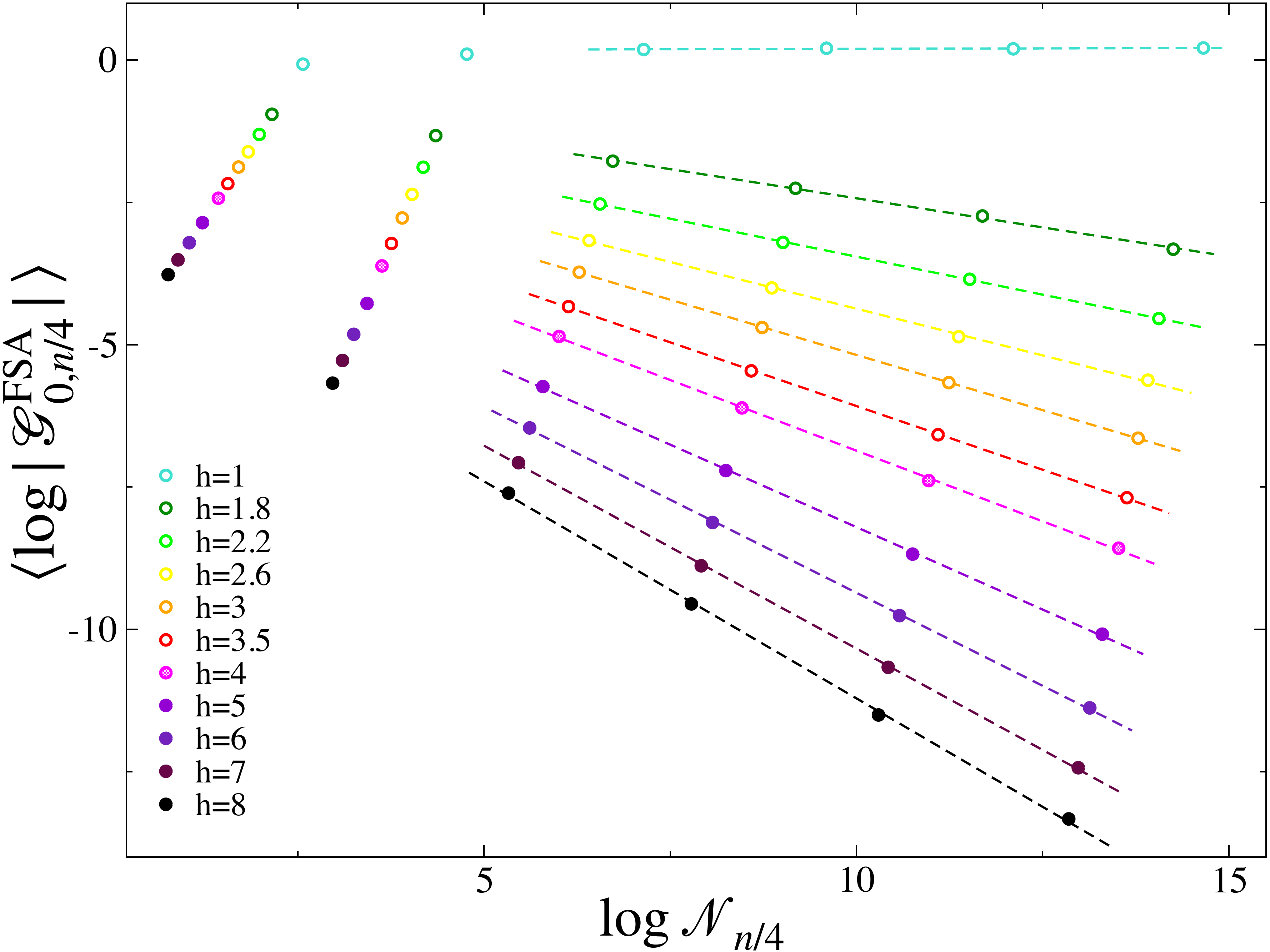}
\caption{\label{fig:FSA}
Logarithm of the typical value of the propagator ${\cal G}_{0,n/4}^{\rm FSA}$ (computed within the FSA) as a function of the (log of the) number of states ${\cal N}_{n/4}$ in the HS at distance $n/4$ from $| 0 \rangle$  and in the same energy shell. Different colors correspond to different values of the disorder across the MBL transition, that previous works estimated in the interval $h_c \in [3.7,4.5]$~\cite{alet,devakul} for chains of about the same length as the ones considered here. The dashed lines correspond to linear fits of the numerical results of the form of Eq.~(\ref{eq:Gtyp}) at large $n$. Filled circles correspond to values of the disorder strength such that $\gamma>1$ ($h>h_c$), while empty circles to $\gamma <1$ ($h<h_c$).}
\end{figure}

\begin{figure}
\includegraphics[width=0.468\textwidth]{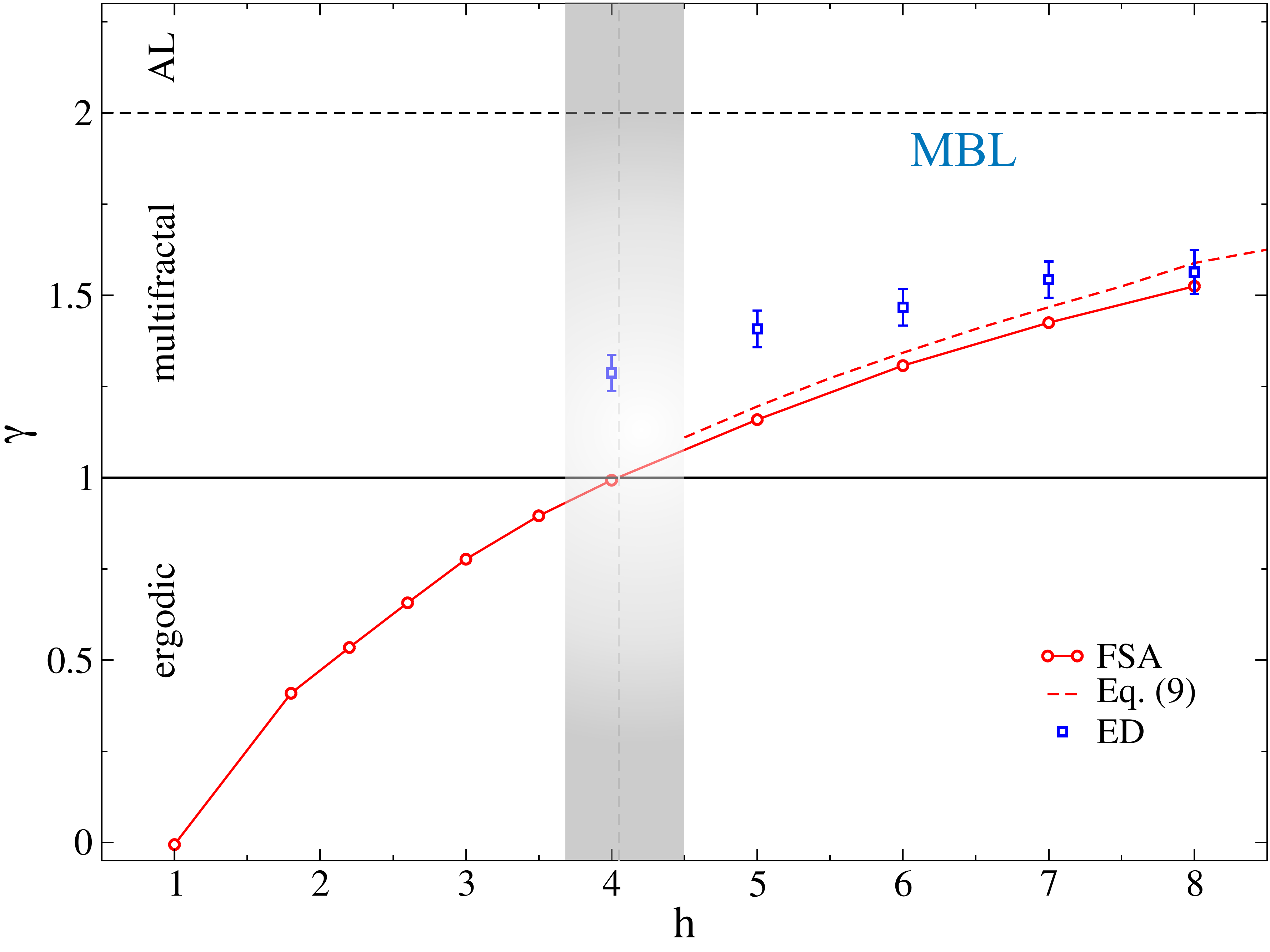}
\caption{\label{fig:gamma}
Effective exponent $\gamma$ describing the scaling of the typical value of the tunneling rates ${\cal G}_{0,n/4}^{\rm FSA}$ between two many-body states which differs by $n/2$ spins with the number of configurations ${\cal N}_{n/4}$ at Hamming distance $n/4$ from a given configuration in the HS, Eq.~(\ref{eq:Gtyp}).
Red circles correspond to the results obtained by linear fitting of $\langle \log | {\cal G}_{0,n/4}^{\rm FSA} | \rangle$ {\it vs} ${\cal N}_{n/4}$ at large $n$, 
Fig.~\ref{fig:FSA}.
The red dashed line corresponds to the strong disorder approximation for $\gamma$ given in Eq.~(\ref{eq:strong_disorder}). The blue squares give the estimation of $\gamma$ obtained by inspecting the unusual scaling of the LDoS computed via exact diagonalizations of~(\ref{eq:Hxxz}), see Sec.~\ref{sec:ldos}.
The gray vertical shaded region marks the disorder range in which the MBL transition is expected to occur according to the most recent numerical results (for chains of about the same size of the ones considered here)~\cite{alet,devakul}, showing that  $\gamma_c = 1$ falls perfectly within it.}
\end{figure}

The sum in~(\ref{eq:FSA}) over the $\textsf{FSP}_{(0,nx)}$ can be efficiently computed exactly using the transfer matrix technique described in~\cite{pietracaprina}. In principle one should estimate the ergodic transition by requiring that {\it any} $x$-sector become ergodic. However for convenience in the following we only focus on the states $\{ {}^{0}\!S_i^z 
 \}$  at zero overlap $q = (1/n) \sum_i {}^{0}\!S_i^z 
 \cdot S_i^z=0$ from the initial one (i.e., when half of the spins have flipped), as schematically depicted in Fig.~\ref{fig:graph}, which is the largest sector on the HS. 
For the XXZ chain the shortest path to achieve $q=0$ (ignoring ``loopy'' terms in which spins are flipped  twice that contribute at higher order in perturbation theory) corresponds to Hamming distance $n/4$ on the graph.
The total number of configurations at distance $n/4$ and from $| 0 \rangle$ and in the same energy shell is ${\cal N}_{n/4} \approx {{n/2}\choose{n/4}}^2 \Omega(E_0) \delta E$, where $\Omega(E_0) = e^{S(E_0)}\propto 1/\sqrt{n}$ is the many-body density of states defined in terms of the microcanonical entropy $S(E_0)$ in the middle of the spectrum.
In Fig.~\ref{fig:FSA} we plot (the log of) the typical value of the matrix elements ${\cal G}_{0,n/4}^{\rm FSA}$ (computed within the FSA) as a function of (the log of) ${\cal N}_{n/4}$  varying the system size from $n=8$ to $n=28$ (the whole probability distributions are shown in Fig.~\ref{fig:PG} of App.~\ref{app}). Different curves correspond to different values of the disorder strength across the MBL transition~\cite{alet,doggen_sub}.
The plot clearly shows that 
\begin{equation} \label{eq:Gtyp}
\left \vert {\cal G}_{0,n/4}^{\rm FSA} \right \vert_{\rm typ}= \mu  \left ( {\cal N}_{n/4} \right)^{-\gamma/2} 
\end{equation}
at large $n$, 
with an exponent $\gamma$ which increases as the disorder is increased (and $\mu$ of order $1$).
The value of $\gamma$ obtained by the linear fitting of $\langle \log | {\cal G}_{0,n/4}^{\rm FSA} | \rangle$ {\it vs} $\log {\cal N}_{n/4}$ (dashed lines of Fig.~\ref{fig:FSA}) is plotted in
Fig.~\ref{fig:gamma} as a function of the disorder $h$, showing that 
the disorder strength at which $\gamma$ becomes larger than one ($h_c \approx 4.05$) happens to be strikingly close to 
the critical disorder of the MBL transition determined by the most recent numerical works (for chains of about the same length of the ones considered here), $h_c \in [3.7,4.5]$~\cite{alet,devakul}. Moreover, we find that in the limit of infinite disorder $\gamma$ tends to $2$. This exact same behavior is found for all the three models considered, as shown in Figs.~\ref{fig:FSAIQP} and~\ref{fig:gammaIQP} of App.~\ref{app}.
The most intuitive way to interpret and rationalize these observations is by using an analogy with a very simple random matrix model, the RP model~\cite{rosenzweig,kravtsov,facoetti} and its generalizations~\cite{nosov,kravtsov1}, which we detail in the next section.

\section{Analogy with the RP model} \label{sec:RP}
The Hamiltonian of the RP model is a $N \times N$ symmetric matrix of the form~\cite{rosenzweig,kravtsov,facoetti}:
\begin{equation} \label{eq:RP}
{\cal H} = {\cal A} + \mu \frac{{\cal M}}{N^{\gamma/2}} \, ,
\end{equation}
where ${\cal A} = a_i \delta_{ij}$ is a diagonal matrix with iid entries drawn from some probability distribution $p(a)$ (whose specific form is unimportant), ${\cal M}$ belongs to the Gaussian Orthogonal Ensemble (GOE) with unit variance (e.g., $m_{ij}$ are iid random Gaussian variables with $\langle m_{ij} \rangle = 0$ and $\langle m_{ij}^2 \rangle = 1$), and $\mu$ is a constant of $O(1)$. 
The first term can be thought as the diagonal quenched on-site disorder, while ${\cal M}$ plays the role of the hopping that might create resonances between two (Poisson distributed) levels close in energy if $|a_i - a_j| \lesssim |m_{ij}|$. The phase diagram of the RP model contains three different phases:
For $\gamma > 2$, standard second-order perturbation theory shows that the GOE term is a small regular perturbation, 
the Hamiltonian is close to ${\cal A}$, and eigenstates are completely localized; Conversely, for $\gamma < 1$ the first term is a small regular perturbation, hence the rotationally invariant term ${\cal M}$  dominates, the eigenstates are uniformly distributed on the unitary sphere, and are fully delocalized;
The regime $1 < \gamma < 2$ is instead special, and provides an example of an extended nonergodic phase~\cite{kravtsov,facoetti}: The eigenstates are supported over a large number of sites---hence they are delocalized---but only over a fraction $N^{1 - \gamma}$ of them ($D_2 = 2 - \gamma$) which tends to zero in the thermodynamic limit---i.e., they are multifractal.

The transition taking place at $\gamma_{\rm AL} = 2$ corresponds to the standard AL and occurs when the amplitude $\Gamma_{\rm AL} = N \langle | {\cal H}_{ij} | \rangle$ vanishes in the thermodynamic limit. The physical interpretation of this criterion is that localization occurs when the number of sites $j$ in resonance with a given site $i$ stays {\it finite} for $N \to \infty$.
The transition from multifractal to ergodic eigenfunctions at $\gamma_c = 1$, instead, occurs when the amplitude
$\Gamma_{\rm ergo} = N \langle {\cal H}_{ij}^2\rangle$
diverges. 
This sufficient criterion for ergodicity has been proposed in Refs.~\cite{nosov,bogomolny,kravtsov1} based on the idea that, using the FGR, the width $\Gamma_{\rm ergo}$ essentially quantifies the escape rate of a particle created in $i$
(note however that the perturbative estimation of $\Gamma_{\rm ergo}$ is valid as long as one can neglect the contribution of the off-diagonal elements to the density of states, i.e., $\gamma > 1$).
For $\gamma<1$ the width $\Gamma_{\rm ergo}$ is much larger than the spreading of energy levels due to the disorder and the system is fully delocalized. For $1 < \gamma < 2$ instead, $\Gamma_{\rm ergo}$ vanishes as $N^{1 - \gamma}$ in the thermodynamic limit, implying that eigenstates only occupy $N^{2 - \gamma}$ sites close in energy.

Thus, adopting this analogy 
and using $\Gamma_{\rm ergo} \to 0$ as the criterion for ergodicity breaking based on the FGR, the FSA estimation of the effective exponents $\gamma$ for the scaling of matrix elements between states at large distance in the HS, 
Eq.~(\ref{eq:Gtyp}) and Fig.~\ref{fig:FSA}), suggests that the MBL phase is similar to the intermediate phase of the RP model~\cite{kravtsov} (and its generalizations~\cite{nosov,kravtsov1}) where $\Gamma_{\rm AL} \to \infty$ but $\Gamma_{\rm ergo} \to 0$, corresponding to delocalized but multifractal eigenstates: 
The initial state $|0 \rangle$ only hybridize with a sub-extensive fraction  
$[ {\cal V} e^{S(E_0)}]^{1 - \gamma}$ of the total configurations, 
thereby producing nonergodic multifractal wavefunctions which only occupy $[{\cal V}e^{S(E_0)}]^{2 - \gamma}$ states close in energy ($D_2 \sim 2 - \gamma$)~\cite{disclaimer}.
In other words, the initial state $|0 \rangle$ is in resonances with {\it many} other states at large distance in the configuration space, but their number is not enough for the quantum dynamics to decorrelate in a finite time.
According to this interpretation, MBL in the HS {\it does not} correspond to AL, which instead would occur for $\gamma >  2$, when the number of resonances found from $| 0 \rangle$ stays {\it finite} in the thermodynamic limit~\cite{bogomolny,nosov,kravtsov1}.

\section{AL on the Bethe lattice within the FSA} \label{sec:Bethe}
In order to elucidate the difference between the criterion $\Gamma_{\rm ergo} \to 0$ proposed above to detect MBL in the HS and standard AL in the limit of infinite dimensions, it is instructive to recall the paradigmatic case of the Anderson tight-binding model on the Bethe lattice~\cite{abou}, described by the Hamiltonian~(\ref{eq:H1}) with $E_a$ iid in the interval $[-W/2,W/2]$. 
As for the MBL case, we determine the probability distribution of the propagator ${\cal G}_{0,n}^{\rm FSA}$ between a point $0$ and one particular point at distance $n$ from it within the FSA~\cite{abou,pietracaprina,dot}.
Let us consider a Bethe lattice of branching number $k$ (total connectivity $k+1$). Without loss of generality the site $0$ can be considered as the root of the tree, and its energy can be set to $E_0 = 0$, in the middle of the band~\cite{disclaimer1}.
Differently from the many-body problem, at the lowest order in perturbation theory there is only one path connecting two sites at distance $n$ on the tree. According to Eq.~(\ref{eq:FSA}) one thus has that ${\cal G}_{0,n}^{\rm FSA} = \prod_{a=1}^n (t/E_a)$. One then immediately finds that on each single path the typical value of the matrix elements decays exponentially as $\langle \log |{\cal G}_{0,n}^{\rm FSA}| \rangle = - n /\xi_{\rm typ}$, over a typical length $\xi_{\rm typ}^{-1} = - \log(2 e t/W)$. Since the number of sites at distance $n$ from $0$ (and in the same energy window) grows as ${\cal N}_n \approx k^n \rho(0) \delta E$, one obtains that $|{\cal G}_{0,n}^{\rm FSA}|_{\rm typ} \propto ({\cal N}_n)^{-\gamma/2}$, with an effective exponent $\gamma = 2/(\xi_{\rm typ} \log k)$. Adopting the ergodicity breaking criterion $\gamma_c = 1$~\cite{nosov,bogomolny,kravtsov1}, we get:
\begin{equation} \label{eq:Wergo}
W_{\rm ergo}^{\rm FSA} = 2 e t \sqrt{k} 
\end{equation}
($W_{\rm ergo}^{\rm FSA}/t \approx 7.7$ for $k=2$). However, such typical decay of the amplitude has nothing to do with Anderson localization, which is instead determined by the requirement that a particle created in $0$ can escape on {\it at least one} of the $k^n$ paths. The localization transition is thus obtained from the decay rate of the maximum amplitude among an exponential number of paths, which for large $n$ is determined by the power law tails of the distribution of ${\cal G}_{0,n}^{\rm FSA}$. This calculation yields the familiar result~\cite{abou,pietracaprina,dot}
\begin{equation}  \label{eq:WAL}
W_{\rm AL}^{\rm FSA} \approx 2 e t k \log k
\end{equation}
($W_{\rm AL}^{\rm FSA} / t \approx 29.1$ for $k=2$, providing un upper bound for the true critical value $W_{\rm AL} \approx 18.2$~\cite{tikhonov_critical}), with a diverging localization length at the transition (with an exponent $\nu = 1$), 
$\xi_{\rm loc} = 1/[2 \log (W/W_{\rm AL})] \gg \xi_{\rm typ}$. 

We argue that the threshold $W_{\rm ergo}^{\rm FSA}$ signals the transition from ergodic to multifractal wavefunctions, which is a genuine phase transition ($W_{\rm ergo} \approx 6.65$ for $k=2$~\cite{DPRM}) only on (loop-less) Cayley trees~\cite{mirlinCT,DPRM,Ioffe,MG} (and is related to the freezing glass transition of directed polymers in random media) and becomes a smooth crossover on the so-called random-regular graphs without boundaries (and loops whose typical size scales as the logarithm of the total number of sites of the graph), where full ergodicity is restored by loops larger than a characteristic correlation length which diverges at $W_{\rm AL}$~\cite{mirlin,gabriel,Bethe}.

Note that in the limit of large connectivity, which is the relevant one for the many-body problem, $W_{\rm ergo}$ and $W_{\rm AL}$ have very different scaling with $k$, as $\sqrt{k}$ and $k \log k$ respectively.
For a disordered XXZ chain of $n$ interacting spins, the average connectivity of the HS grows as $n/2$, the effective width of the disorder is of order $h \sqrt{n/12}$, and $t=1/2$.
Using the Bethe lattice estimation for $W_{\rm ergo}^{\rm FSA}$, Eq.~(\ref{eq:Wergo}), one obtains a critical disorder $h_c \approx e \sqrt{6}  \approx 6.7$.
The scaling of $W_{\rm AL}$ with the connectivity, Eq.~(\ref{eq:WAL}), would instead prohibit AL (note however the strong correlations of the potential between neighboring sites~\cite{logan}).

Assuming that in the Hilbert space the typical value of the propagator decays exponentially with the distance over a typical length scale, $|{\cal G}_{0,nx} |_{\rm typ} \propto e^{-n x / \xi_{\rm typ}}$, since the number of nodes in the Hilbert space at Hamming distance $n/4$ from $|0\rangle$ is ${\cal N}_{n/4} = {{n/2}\choose{n/4}}^2 e^{S(E_0)} \sim 2^n/n^{3/2}$, the amplitude 
	$\Gamma_{\rm ergo}$ can be expressed as $\Gamma_{\rm ergo} = |{\cal G}_{0,n/4} |_{\rm typ}^2 \, {\cal N}_{n/4} \sim e^{- n / (2 \xi_{\rm typ}) + n \log (2) + O(\log n)}$. Hence, full delocalization and ergodicity occur when $\xi_{\rm typ} \ge \xi_{\rm typ}^c =  1/(2 \log 2)$. This is in a certain sense the analogue of  the sufficient condition of delocalization obtained in~\cite{aizenman} for the Anderson model on the BL, which correspond to the requirement that the exponential decay of  typical  correlations does not compensate anymore for the exponential proliferation of sites at large distance. Surprisingly enough, the  existence of such {\it universal} value of  $\xi_{\rm typ}^c$ at the MBL transition predicted by   this simple argument is in perfect agreement with the recent results of the phenomenological RG approach of Ref.~\cite{KT1}.

This simple example clearly illustrates the differences that arise between AL on the Bethe lattice and MBL in the HS (at least within the FSA).
As mentioned above, random on-site effective energies of the many-body problem are not independent variables and are strongly correlated~\cite{logan}. Moreover, 
the  number of forward-scattering paths connecting two many-body configurations grows factorially with the length of the paths, while on the Bethe lattice two points are connected by a unique path at the lowest order in $t/W$. However our analysis indicates that probably the most important difference consists in the fact that while AL on the Bethe lattice occurs when the number of resonances  found at large distance from a given site are {\it finite} in the thermodynamic limit (i.e., single-particle eigenstates occupy a {\it finite} volume in the thermodynamic limit), 
MBL instead takes place when the number of resonances found at distance $nx$ from a given many-body state is still large but not enough to ensure e ergodicity and to allow the quantum dynamics to decorrelate from $|0\rangle$ in a finite time (i.e., many-body eigenstates are {\it extended but non-ergodic} and occupy a sub-extensive portion $[{\cal V}e^{S(E_0)}]^{D_2}$ of the  accessible volume in the HS). 
Hence, while single-particle AL on the BL is governed by the {\it tails} of the distribution $P({\cal G}_{0,n})$, MBL in the HS is governed by its {\it bulk} properties. 
Of course, focusing only on the typical value of the matrix elements, as done in Sec.~\ref{sec:FSA} is a drastic, and possibly wrong, assumption, since it neglects the effect of strong rare resonances that are known to play a very important role in MBL~\cite{avalanches,gopala} (and probably increases the estimate for the critical disorder $h_c$~\cite{doggen_sub}). We will come back to this issue in Sec.~\ref{sec:2d} and in the concluding section, arguing that it might be corrected by mapping the MBL problem in the HS onto suitable generalizations of the RP model with power-law distributed off-diagonal elements~\cite{kravtsov1}, and possibly refining the FSA computation by adding the self-energy corrections~\cite{logan1,bogomolny} in the denominators of~(\ref{eq:FSA}) and/or higher order terms of the locator expansion~\cite{anderson,colmenarez} in order to describe more accurately the tails of the distribution $P({\cal G}_{0,nx})$.

\section{Strong disorder approximation} \label{sec:SD}
In this section we discuss how the effective exponent $\gamma$ describing the scaling of the typical tunneling rates between two many-body states separated by $nx$ spin flips with the number of configurations ${\cal N}_{n x}$ at distance $nx$ from a given configuration in the HS, Eq.~(\ref{eq:Gtyp}), can be estimated analytically in the limit of strong disorder within the FSA.

A first very naive estimation 
might be obtained by recalling the intuitive argument given in the introduction for the origin of the multifractality of the many-body eigenstates in the HS, due to the presence of a finite density of active spins on the sites where the random fields is smaller than the energy required for spin flips~\cite{mace,laflorencie,resonances}. For the disordered XXZ spin chain the energy needed to flip two neighboring spins with opposite sign, e.g. $i$ and $i+1$, is $\Delta E = \pm (h_i - h_{i-1}) + m \Delta /2$, with $m=0,+1,-1$ depending on how many domain walls 
have been created (annihilated) in the process. In the limit $h \gg \Delta$ the second term can be neglected, and $\Delta E$ is then a random variable of zero mean and variance $2 h^2/3$. The density of active pairs of spins is thus proportional to the probability that this random variable is smaller than $t=1/2$, $\rho \sim \sqrt{3}/(2 \sqrt{2} h)$
Since many-body states have typically $n/2$ domain walls, the volume occupied by many-body wavefunctions in this limit is $2^{\frac{n}{2} \rho}$. Assuming, by analogy with the RP model, that $D_2 \sim \rho/2 \sim 2 - \gamma$~\cite{kravtsov}, one obtains that 
\[
\gamma \sim 2 - \frac{\sqrt{3}}{4 \sqrt{2} h}  \, .
\]
However this expression gives a very poor approximation of the numerical results plotted in Fig.~\ref{fig:gamma}, and overestimates $\gamma$ by a large amount due to the fact that resonances and hybridization beyond the nearest neighboring spins are completely neglected.

A slightly more refined calculation can be performed as described below.
Let us consider a random initial state $| 0 \rangle = | \! \! \uparrow \downarrow \uparrow \cdots \rangle$ with energy $E_0$ in the middle of the many-body spectrum.
In order to reach a state at zero overlap from it 
at the lowest order in the hopping we have to flip $n/4$ pair of spins in {\it different} locations of the chain.
Consider one particular path joining $| 0 \rangle$ with a given state at distance $n/4$ from it. The on-site random energies on the states visited along the path evolve as $E_0 \to E_0 + \Delta E_1 \to E_0 + \Delta E_1 + \Delta E_2 \to \ldots E_0 + \sum_{i =1}^{n/4} \Delta E_i$ (recall that for $\ell$ large $\sum_{i =1}^{\ell} \Delta E_i$ is a Gaussian random variable of zero mean and variance $2 \ell h^2 /3$). The contribution to the sum~(\ref{eq:FSA}) coming from this specific path is then:
\begin{equation} \label{eq:1path}
\frac{1}{2 \Delta E_1} \times \frac{1}{2(\Delta E_1 + \Delta E_2)} \times \cdots \times \frac{1}{2 \sum_{i =1}^{n/4} \Delta E_i} \, ,
\end{equation}
which is a random variable.
All the $(n/4)!$ permutations of the sequence by which spins are flipped yield a different path contributing to the same matrix elements between the same two states. 
In the following we assume that the contributions of different paths are uncorrelated and that the variance of the product~(\ref{eq:1path}) on a single path is finite. Both assumptions are of course wrong. Yet, since we are only interested in the typical scaling of the matrix elements, they might give a reasonable approximation for the effective exponent $\gamma$.
The total number of configurations in the HS at distance $n/4$ from  $| 0 \rangle$ and in the same energy shell is ${\cal N}_{n/4} \propto {{n/2}\choose{n/4}}^2 \Omega(E_0)$, with $\Omega(E_0) = e^{S(E_0)} \propto 1/\sqrt{n}$ being the many-body density of states in the middle of the many-body spectrum (and $S(E_0)$ is the microcanonical entropy).
An approximate estimation of the exponent $\gamma$ defined in Eq.~(\ref{eq:Gtyp}) is then given by:
\begin{equation} \label{eq:strong_disorder}
\gamma \underrel{n\to \infty}{\approx} \frac{\frac{1}{2}\log \frac{n}{4}! - \langle \log(2 \Delta E_1)\rangle  - \cdots - \langle \log(2 \sum_{i=1}^{n/4} \Delta E_i)\rangle}{2 \log {{n/2}\choose{n/4}} - \frac{1}{2} \log n} \, .
\end{equation}
The prediction of Eq.~(\ref{eq:strong_disorder}) is plotted in Fig.~\ref{fig:gamma}, showing a good agreement with the numerical results obtained by linear fitting of the data of Fig.~\ref{fig:FSA} at large $n$ in the whole range $[h_c,\infty)$ (similar results are found also for the Imbrie model, Fig.~\ref{fig:gammaIQP} of App.~\ref{app}).

\begin{figure}
\includegraphics[width=0.468\textwidth]{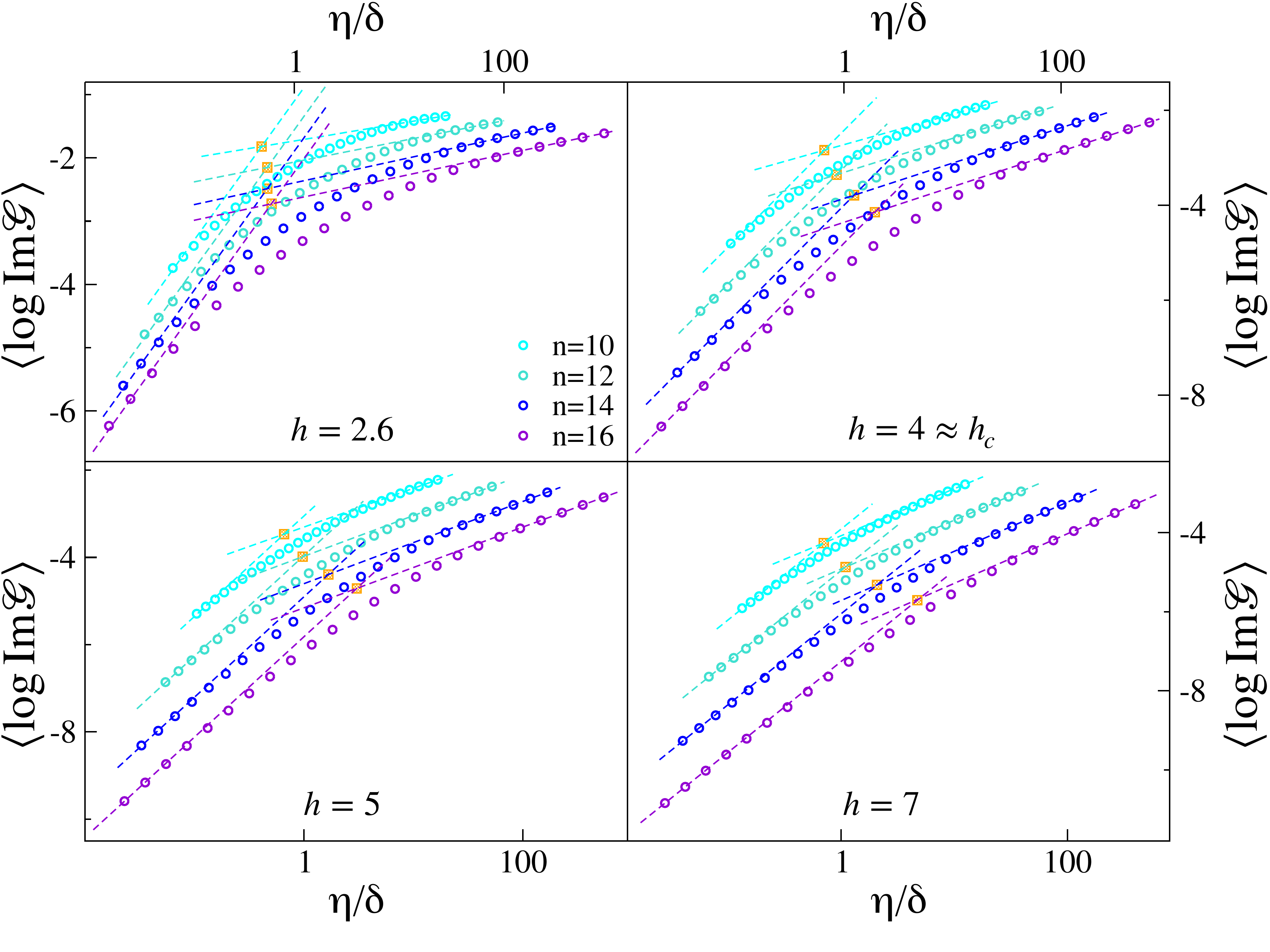}
\caption{\label{fig:ldos}
$\langle \log {\rm Im}{\cal G} (E_0 + i \eta) \rangle$ as a function of the imaginary regulator expressed in units of the mean level spacing, $\eta/\delta$, for four different values of the disorder across the MBL transition, $h=2.6$ (top-left panel), $h=4 \approx h_c$ (top-right panel), $h=5$ (bottom-left panel), and $h=7$ (bottom-right panel), and for four chain lengths, $n=10$ (cyan), $n=12$ (turquoise), $n=14$ (blue), and $n=16$ (violet).
The LDoS is obtained from Eq.~(\ref{eq:ldos}) by inverting exactly the many-body Hamiltonian~(\ref{eq:Hxxz}) in the spin configuration basis, and averaging over $2^{40-2n}$ independent realizations of the disorder. The dashed straight lines represents the linear fits of $\langle \log {\rm Im}{\cal G} (E_0 + i \eta) \rangle$ as a function of $\log (\eta / \delta)$ at small and large $\eta$ with slope $1$ and $\zeta(h)$ respectively, Eq.~(\ref{eq:fitldos}). The orange squares mark the crossing points of the two straight lines which yield our estimation of the crossover scale $\eta_\star$.}
\end{figure}

\section{Statistics of the local density of states} \label{sec:ldos}
In this section we backup the perturbative results obtained within the FSA by probing the non-standard scaling limit of the local density of states (LDoS) which gives direct access to the nonergodic features of wavefunction statistics and provides an independent estimation of the fractal spectral dimension $D_1$.

The LDoS computed on a particular ``site orbital'' $|a \rangle$ of the HS is:
\begin{equation} \label{eq:ldos}
{\rm Im}{\cal G}_{aa} (E_0 + i \eta) = \sum_{\beta=1}^{\cal V} | \psi_\beta (a) |^2 \frac{\eta}{(E_\beta - E_0)^2 + \eta^2} \, ,
\end{equation}
where $|a \rangle = | \! \uparrow \downarrow \uparrow \cdots \rangle$, $E_0 =  -n/(4n-4)$ is the energy in the middle of the many-body spectrum, ${\cal V} = {{n}\choose{n/2}}$ is the dimension of the HS, $E_\beta$ are the eigenvalues of the Hamiltonian~(\ref{eq:Hxxz}), $\psi_\beta (a)$ are the eigenfunctions' amplitudes expressed in the spin configuration basis, and $\eta$ is an additional imaginary regulator.
The average value of this quantity gives the many-body density of states $(1/\pi) \langle {\rm Tr} {\rm Im} {\cal G} (E_0) \rangle = e^{S(E_0)}$. 
In contrast, the typical value of ${\rm Im}{\cal G}$ is controlled by the matrix element that couples a given site to the resonance sites at energy $E_0$.
Let us imagine a situation in which the sum in~(\ref{eq:ldos}) contains only $[{\cal V} e^{S(E_0)}]^{D_1}$ peaks of significant weight. Then the LDoS becomes smooth only if the broadening $\eta$ exceeds the typical spacing between the peaks which, in the middle of the many-body spectrum, is typically of the order $[{\cal V} /\sqrt{n}]^{-D_1}$.
This implies that the typical value of the LDoS should exhibit the characteristic localized behavior (i.e., $\langle \log {\rm Im}{\cal G} \rangle \propto \eta$) up to a characteristic crossover scale $\eta_\star$ much larger than the mean level spacing $\delta \approx \sqrt{n}/{\cal V}$.

\begin{figure}
\includegraphics[width=0.49\textwidth]{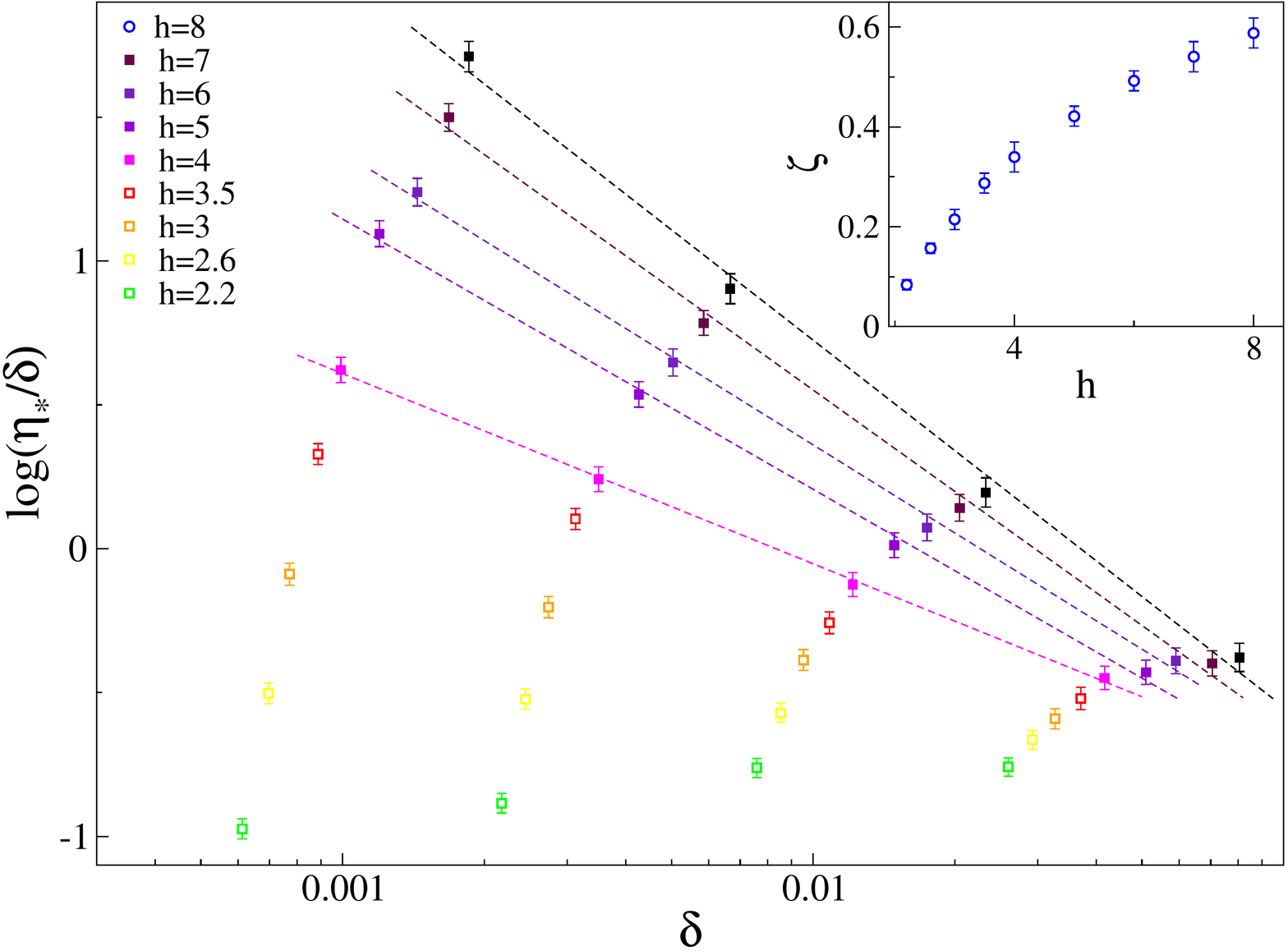}
\caption{\label{fig:etastar}
Main panel: Logarithm of the crossover scale $\eta_\star / \delta$, extracted from the plots of Fig.~\ref{fig:ldos} as explained in the main text, as a function of the mean level spacing $\delta$ for several values of the disorder strength $h$ and several system sizes $n$ from $10$ to $16$. The dashed line correspond to fits of the data of the form $\eta_\star / \delta \propto \delta^{D_1 - 1}$ for $h>h_c$, which gives an estimation of the effective exponent $\gamma = 2 - D_1$~\cite{kravtsov} (blue squares of Fig.~\ref{fig:gamma}). Inset: Exponent $\zeta(h)$ describing the behavior of the typical value of the LDoS for $\eta \gg \eta_\star$ as a function of the disorder strength.}
\end{figure}

In Fig.~\ref{fig:ldos} we plot the logarithm of the typical value of the LDoS, $\langle \log {\rm Im}{\cal G} \rangle$, as a function of the imaginary regulator measured in units of the mean level spacing, $\eta/\delta$, for several system sizes ($n$ from $10$ to $16$) and for four values of the disorder strength $h$ across the MBL transition. These plots are obtained by inverting exactly the many-body Hamiltonian~(\ref{eq:Hxxz}) in presence of the imaginary regulator, and averaging over several (about $2^{40-2n}$) independent realizations of the disorder. The curves clearly show the existence of the crossover scale $\eta_\star$ such that
\begin{equation} \label{eq:fitldos}
\left \{
\begin{array}{ll}
{\rm Im} {\cal G}_{\rm typ} \propto \eta & \textrm{for~} \eta \ll \eta_\star \\
{\rm Im} {\cal G}_{\rm typ} \propto \eta^{\zeta(h)} & \textrm{for~} \eta \gg \eta_\star 
\end{array}
\right .
\end{equation}
with an exponent $\zeta(h)>0$ (except at very small $h$) which depends on the disorder (see inset of Fig.~\ref{fig:etastar}) but not on the system size. [Note that for completely AL eigenstates one should observe instead ${\rm Im} {\cal G}_{\rm typ} \propto \eta$ up to $\eta$ of $O(1)$.] Concretely, we have measured $\eta_\star$ by performing linear fits of $\langle \log {\rm Im} {\cal G} \rangle$ at small and large $\eta$ with slope $1$ and $\zeta(h)$ respectively, and determining where the two straight lines cross (orange squares of Fig.~\ref{fig:ldos}).
The crossover scale $\eta_\star$ obtained by applying this procedure is plotted in Fig.~\ref{fig:etastar} as a function of the mean level spacing $\delta$ for several values of $h$.
In the MBL phase, $h \ge h_c$, $\log (\eta_\star / \delta)$ increases linearly by decreasing $\log \delta$ (i.e., increasing $n$), consistently with the presence of multifractal eigenfunctions which only  occupy a subextensive part of the HS.
By fitting $\eta_\star / \delta \propto \delta^{D_1 - 1}$ one obtains a 
measure of the fractal dimension $D_1$ which, by analogy with the RP model, gives a rough estimation of the effective exponent $\gamma = 2 - D_1$~\cite{kravtsov} (blue squares in Fig.~\ref{fig:gamma}).
Instead on the metallic side, $h < h_c$, one observes a deviation from the straight line for the largest system sizes, signaling the recovery of a fully ergodic behavior.
In particular at small enough disorder ($h=2.6$, top-left panel of Fig.~\ref{fig:ldos}) we find that $\eta_\star \approx \delta$, implying that $D_1 = 1$, as expected for fully ergodic eigenstates. Similar results are found also for the Imbrie model, Fig.~\ref{fig:ldosI} of App.~\ref{app}.
 
Note that the anomalous scaling limit of the spectral statistics in the multifractal regime $1 < \gamma < 2$ of the RP model 
has been analyzed in full details~\cite{facoetti}, and turns out to be slightly different from the one observed for the MBL system and shown in Fig.~\ref{fig:ldos}. In particular for the RP model one expects a region $\eta_\star <\eta < \eta_{\rm th}$ where ${\rm Im} {\cal G}_{\rm typ}$ is independent of $\eta$ due to the presence of {\it mini-bands} of eigenfunctions close in energy and occupying $N^{D_1}$ sites. The Thoules energy $\eta_{\rm th}$ is the width of these mini-bands and is simply obtained by multiplying the number of levels within a mini-band times the mean level spacing, i.e., $\eta_{\rm th}/\delta = N^{D_1}$. Hence the typical value of the LDoS, when plotted as a function of $\eta / \delta$ should exhibit a flat part for $N^{D_1-1} \ll \eta/\delta \ll N^{D_1}$ which becomes broader and broader as the system size is increased.
The absence of such flat region in Fig.~\ref{fig:ldos} indicates that, differently from the RP model, for the many-body Hamiltonian~(\ref{eq:Hxxz}) the mini-band in the LDoS are multifractal~\cite{QREM}.  

\section{ Limitations of the FSA and the example of two-dimensional systems} \label{sec:2d}

The main problem of the approach put forward in this work comes from the fact that the non-renormalized perturbative expansion has poles for any value of the energy within the support of the probability distribution of the random energies. In fact the non-renormalized perturbative expansion of the resolvent is always (i.e. with probability $1$ in the thermodynamic limit) divergent even in the localized phase, due to local resonances, i.e. the sites in the expansion~(\ref{eq:FSA})  where $|E-E_c|<t$, whose presence is inevitable in the thermodynamic limit however strong the disorder might be. Yet physically this is not a problem for localization. In fact the exact poles  of the Green's functions should be found at the eigenenergies of the Hamiltonian and not at the random on-site energies. As shown by Anderson~\cite{anderson}, this issue should be solved by re-summing the closed paths in the series expansion through the self-energy corrections. Since this re-summation cannot be done exactly on a generic lattice, in this work we have chosen to retain only the leading order terms of the perturbative expansion and to focus only on the scaling of the {\it typical} value of the propagator, which is only weakly affected by the presence of the poles.
Nonetheless, by doing so we are possibly overlooking  the effect of rare large amplitudes of the tunneling rates in the tails of the distribution (see e.g. Fig.~\ref{fig:PG} of App.~\ref{app}), 
whereas  rare delocalizing process (also called “thermal inclusions”) are known to play a  crucial role in the context of MBL.
In fact, in the latest years a phenomenological description for the many-body delocalization was proposed~\cite{avalanches}, which relies on the “avalanche” instability, i.e., proliferation of an initial effectively thermal seed which grows until it swallows the whole system for $h<h_c$~\cite{thiery,avalanches,gopala}.
The  avalanche picture predicts for instance that thermalization avalanches should destabilize the MBL phase in any dimension larger than $1$~\cite{gopala}.

On the one hand, the ergodicity breaking criterion obtained from the typical decay of the matrix elements, ${\cal N}_{n x} |{\cal G}_{0,nx}|_{\rm typ}^2 \to 0 $, 
yields a critical disorder which is in strikingly good agreement with the most recent numerical results obtained from exact diagonalizations of chains of about the same length of the ones considered here~\cite{alet,devakul,abanin,roscilde}.
On the other hand, however,  
recent results obtained in by applying time-dependent variational principle to matrix product states that allow to study $1d$ chains of a length up to $n=100$ indicate a substantial increase of the estimate for the critical disorder ($h_c \gtrsim 5.5$) that separates the ergodic and many-body localized regimes~\cite{doggen_sub}. Such enhancement of ergodicity in large systems is likely to be due to  the  existence  of rare 
delocalizing processes~\cite{avalanches,gopala}---possibly involving degrees of freedom distant in real space---that are not typically present in small systems and that are not detectable by typical amplitudes.

\begin{figure}
 \includegraphics[width=0.54\textwidth]{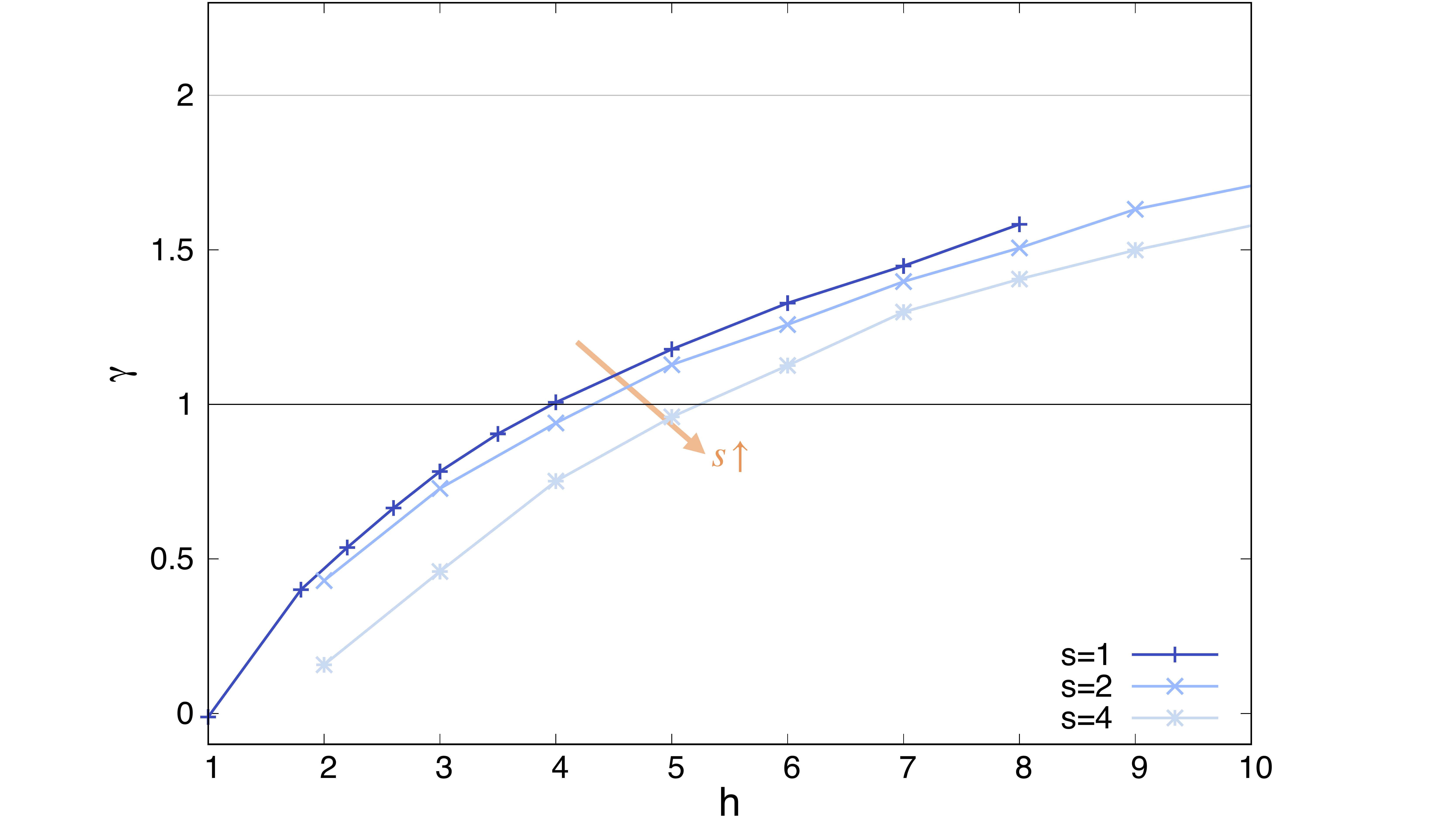}
	\caption{\label{fig:2d} Effective exponent $\gamma$ describing the scaling of the typical value of the matrix elements ${\cal G}_{0,n/4}^{\rm FSA}$ between two many-body states at Hamming distance $n/4$ in the HS (i.e., such that half of the bosons have moved compared to the initial state) for the model~(\ref{eq:HCB}) varying the width of the strip, $s=1$ (same curve as in Fig.~\ref{fig:gamma}), $s=2$ and $s=4$.}
\end{figure}

In order to have a concrete benchmark example of the possible limitations of our approach, in this section we apply the ideas discussed above  to a two-dimensional setting, where thermalization avalanches are expected to destabilize the insulating phase in the thermodynamic limit. More specifically, 
we consider hard-core bosons with nearest-neighbor interactions on a quasi-$1d$ square lattice of length $L$ and width $s$ and total number of sites $n=Ls$~\cite{doggen}. The Hamiltonian is given by:
\begin{equation} \label{eq:HCB}
{\cal H}_ {\rm HCB} = \sum_{\langle i, j \rangle} \left[ - \frac{J}{2} \left( b_i^\dagger b_j + \textrm{h.c.} \right) + U n_i n_j \right] - \sum_i \epsilon_i n_i \, ,
\end{equation}
where $b^\dagger_i$ creates a boson at site $i$, $n_i  = b^\dagger_i b_i$ (the occupation of each site is restricted to $n_i = 0,1$), and 
the summation over $\langle i, j \rangle$ couples neighboring sites on the strip. The on-site potentials $\epsilon_i$ are iid random variables taken from a uniform distribution on $[-h,h ]$. We set $J=1$ and $U=1$ in the half-filling sector, $\sum_i n_i = n/2$ .

For $s=1$ this model is equivalent via a Jordan-Wigner transformation to the Heisenberg XXZ random-field spin chain~(\ref{eq:Hxxz}) considered above, while upon increasing the width $s$ of the strip we move towards a $2d$ geometry. Below we repeat the same analysis described in Sec.~\ref{sec:FSA} for this model with $s=2$ and $s=4$. 
We choose as a basis in the HS the tensor product of the simultaneous eigenstates of the number operators $n_i$, 
$| a \rangle = | \!\bullet \! \circ \! \bullet \! \cdots \rangle$, 
such that for $J=0$ the many-body eigenstates are perfectly localized.
We pick an infinite temperature many-body state $| 0 \rangle$ at random with energy close to the middle of the many-body
spectrum, and we evaluated the effective tunneling rates between such state and states at  Hamming distance $n/4$ in the HS, i.e., such that half of the bosons have moved compared to the initial configuration. The computation of the propagator  is again performed at the level of the FSA, i.e., retaining only the $\sim (n/4)!$ leading-order terms in the perturbative expansion (i.e., only the shortest paths in the HS).
By comparing the scaling of the typical value of the matrix elements $|{\cal G}^{\rm FSA}_{0,n/4}|_{\rm typ}$ with the number of accessible nodes of the HS at Hamming distance $n/4$ from $| 0 \rangle$ and in the same energy shell, ${\cal N}_{n/4} = {{n/2}\choose{n/4}}^2 e^{S(E_0)}$, we obtain the effective exponent $\gamma$ plotted in Fig.~\ref{fig:2d} for $s=1$ (same curve as in Fig.~\ref{fig:gamma}), $s=2$ and $s=4$. 

Applying the  ergodicity breaking criterion ${\cal N}_{n/4} |{\cal G}_{0,n/4}|_{\rm typ}^2 \to 0$ inspired by the RP model~\cite{kravtsov} and its generalizations~\cite{bogomolny,nosov,kravtsov1}, we obtain that the critical disorder $h_c$ where $\gamma$ becomes larger than one does systematically (possibly exponentially) increase with the width of the strip $s$, as predicted by the avalanche approach. Yet, such increase is much weaker than the one recently obtained in~\cite{doggen}  by applying time-dependent variational principle to very large quasi-$1d$ strips, and is also much weaker than the analytical prediction of the effect of  avalanches in $2d$, $h_c \propto 2^s$~\cite{gopala,doggen}. 

It is also instructive to study how the whole probability distribution of the propagator is modified upon increasing $s$ (at the same disorder strength in the MBL regime, $h=6$, and for the same total number of sites $n=Ls =24$), as plotted in Fig.~\ref{fig:PG2d} of App.~\ref{app}. One clearly observes a strong enhancement of the tails of the distribution when $s$ is increased, corresponding to rare large tunneling amplitudes, which, however, have only a moderate effect on the typical value.

All in all, this analysis indicates that the  approach put forward in this work is  able to capture some some mild signature of  the avalanche instability. Yet, focusing only on the typical value of the propagator evaluated at the leading order of the perturbative expansion does not allow to recover the full correct quantitative behavior, especially in situations in which the thermal inclusions are expected to have a strong effect (i.e., very large systems and/or $d>1$). 

\section{Discussion and conclusions} \label{sec:conclusions}
In this paper we have proposed  a novel perspective to analyze the properties of the MBL transition in the HS~\cite{dot,Jacquod,scardicchio_bethe_mbl,dinamica,logan1,yukalov}.
We evaluated the tunneling rates ${\cal G}_{0,nx}^{\rm FSA}$ between two many-body states at extensive distance $nx$ (when a finite fraction $x$ of the spin are flipped) at the lowest order in perturbation theory starting from the insulator, and compared their (typical) amplitude to the number of accessible configurations ${\cal N}_{n x}$ at distance $nx$ from a given initial state (and in the same energy shell). 
We have shown that in the MBL phase, although typically {\it many} resonances are formed, they are not enough to allow the quantum dynamics to decorrelate from the initial condition in a finite time.
Concretely, we have put forward a 
criterion for ergodicity breaking 
based on the ideas of Refs.~\cite{bogomolny,nosov,kravtsov1} and on the FGR. This criterion 
is much weaker than requiring AL of many-body eigenstates in the HS and
suggests that the MBL transition takes place when the amplitude $\Gamma_{\rm ergo} = {\cal N}_{n x} |{\cal G}_{0,nx}|_{\rm typ}^2 \to 0$~\cite{disclaimer}. 
This implies that many-body eigenfunctions in the HS are {\it delocalized} but {\it multifractal}, and typically only occupy a subxtensive portion $[ {\cal V} e^{S(E_0)}]^{D_2}$ of the total accessible configurations, with $0<D_2<1$ and $D_2 \to 0$ only in the limit of infinite disorder. 
MBL in the HS is thus reminiscent of the transition from ergodic to multifractal states of the RP matrix ensemble~\cite{kravtsov} (and its generalizations~\cite{nosov,kravtsov1}), 
and {\it not} to AL in the limit of infinite dimension~\cite{abou}, which instead occurs when the number of resonances found from a given configuration stays {\it finite} in the thermodynamic limit~\cite{bogomolny,nosov,kravtsov1} (in fact we find that many-body eigenstates become AL only in the limit of infinite disorder, as expected from intuitive arguments~\cite{QREM}).

Our interpretation fully supports the picture recently proposed in Ref.~\cite{laflorencie} where MBL is seen as a fragmentation of the HS, as well as similar ideas promoted to explore the analogy between MBL  and a percolation transition in the configuration space~\cite{chalker}.
It is also in agreement with the most recent numerical results on the wavefunctions' statistics~\cite{mace}, with perturbative calculations~\cite{resonances}, and with intuitive arguments that strongly indicate that the many-body eigenstates are multifractal in the whole MBL regime.  

The approach presented in this paper has several advantages. On the one hand, it provides a clear view of the MBL transition in the HS, which is conceptually of prime interest and gives a transparent explanation of the difference between AL and MBL; On the other hand, it yields a simple and quantitatively predictive tool to estimate the critical disorder and the properties of the eigenstates in the MBL phase, since the transfer matrix algorithm~\cite{pietracaprina}  used to determine the amplitude of the propagator~(\ref{eq:FSA}) is computationally much easier than exact diagonalizations and allows to investigate larger system sizes.

In the following we discuss several possible  limitations and implications of our approach, as well as some perspectives for future work.

{\bf (1)}As already discussed in details in Sec.~\ref{sec:2d}, in this work we only focused on the asymptotic scaling behavior of the {\it typical} value of the amplitude of the propagator evaluated at the lowest order of the perturbative expansion. 
In this way we are clearly overlooking the effect of strong rare resonances in the tails of the distributions of the propagators (see Fig.~\ref{fig:PG}). By doing so we find that the ergodicity breaking criterion built on the FGR, $\gamma_c = 1$, yields a critical disorder which is in strikingly good agreement with the most recent numerical results obtained from exact diagonalizations of chains of about the same length of the ones considered here~\cite{alet,devakul,abanin,roscilde}.
However, 
our approach is not able to capture the enhancement of ergodicity observed  in very large chains~\cite{doggen_sub} and in two-dimensional systems~\cite{doggen,gopala} which is likely to be due to  the  existence  of rare 
delocalizing processes~\cite{avalanches,thiery,gopala}---possibly involving degrees of freedom distant in real space---that are not typically present in small systems and that are not detectable by typical amplitudes.
It would be therefore helpful to go beyond the FSA either including higher order terms in the perturbative expansion (see e.g.~\cite{colmenarez} for a recent attempt in this direction), or developing  some approximate treatment for the  self-energy corrections in the denominators of~(\ref{eq:FSA}), as, for instance, recently proposed in Refs.~\cite{logan,bogomolny}. This might  allow one to  describe more accurately the tails of the distribution $P({\cal G}_{0,nx})$ and to  take into account the effect of rare resonances which are known to play a very important role in the context of MBL~\cite{avalanches}.

{\bf (2)} A tightly related issue is that the RP model is certainly oversimplified: The mini-band in the local spectrum is not multifractal, the spectrum of fractal dimension is degenerate, and strong resonances are absent (see above).
In fact, we have already noticed in Sec.~\ref{sec:ldos} that the typical value of the LDoS of the many-body Hamiltonian, Fig.~\ref{fig:ldos}, behaves differently from the one of the RP model~\cite{facoetti}. 
In this respect it would be useful to study suitable generalizations of the RP ensemble with broadly distributed off-diagonal elements (see, e.g.,~\cite{kravtsov1}) that might provide a better effective description for 
the MBL transition in the HS. 
It would be also desirable to complete the present computation by studying the asymptotic scaling of the propagator in all the $x$-sectors, which might  allow one to obtain a more precise estimation of the effective exponent $\gamma$.

{\bf (3)} Another related problem is that the ergodicity breaking criterion used here, together with the mapping onto the RP model, seem to predict that the spectral dimension $D_2 \sim 2 - \gamma$ is continuous at the MBL transition (i.e., $\gamma \to 1$ for $h \to h_c^+$), while recent numerical results~\cite{mace} as well as theoretical arguments~\cite{avalanches} indicate a discontinuous jump of $D_q$ at $h_c$. 
It would be interesting to understand whether going beyond the FSA by including higher order corrections to Eq.~(\ref{eq:FSA}) and/or considering generalizations of the RP ensemble with broadly distributed off-diagonal matrix elements  as effective descriptions of the MBL transition in the HS can lead to a scenario in which the fractal dimensions exhibit a discontinuous jump at the transition. 

{\bf (4)} Another important aspect concerns the implications of the interpretation proposed here on the unusual properties of the bad metal phase preceding MBL~\cite{bad_metal1,bad_metal2}.
In some recent works it was in fact suggested that the subdiffusive transport and the anomalously slow out-of-equilibrium relaxation observed in numerical simulations and experiments might be explained in terms of the apparent nonergodic features of many-body wavefunctions in the HS~\cite{dinamica,DPRM,BarLevnonergo}, while the interpretation proposed here and recent numerical results~\cite{mace} indicates that eigenstates of~(\ref{eq:Hxxz}) become fully ergodic on the metallic side of the transition. This issue might be explained in terms of strong finite-size effects. 
The scaling analysis of~\cite{mace} indicates indeed that for $h<h_c$ the finite-size effects controlling the asymptotic scaling behavior of the inverse participation ratios $\Upsilon_q$ are dominated by a nonergodic volume which diverges very fast as the transition is approached (see also~\cite{mirlin,gabriel,Bethe}). 
As a result, many-body wavefunctions might behave as if they were multifractal in a very broad range of system sizes even before $h_c$, especially in the region in which $\gamma$ is close to $1$ and the spectral band-width associated to the off-diagonal tunneling rates is of the same order of the spreading of the energy levels due to the disorder, thereby producing anomalous diffusion and slow power-law relaxation of physical correlations on a very large time-window spanning many decades.

\begin{figure}
\includegraphics[width=0.49\textwidth]{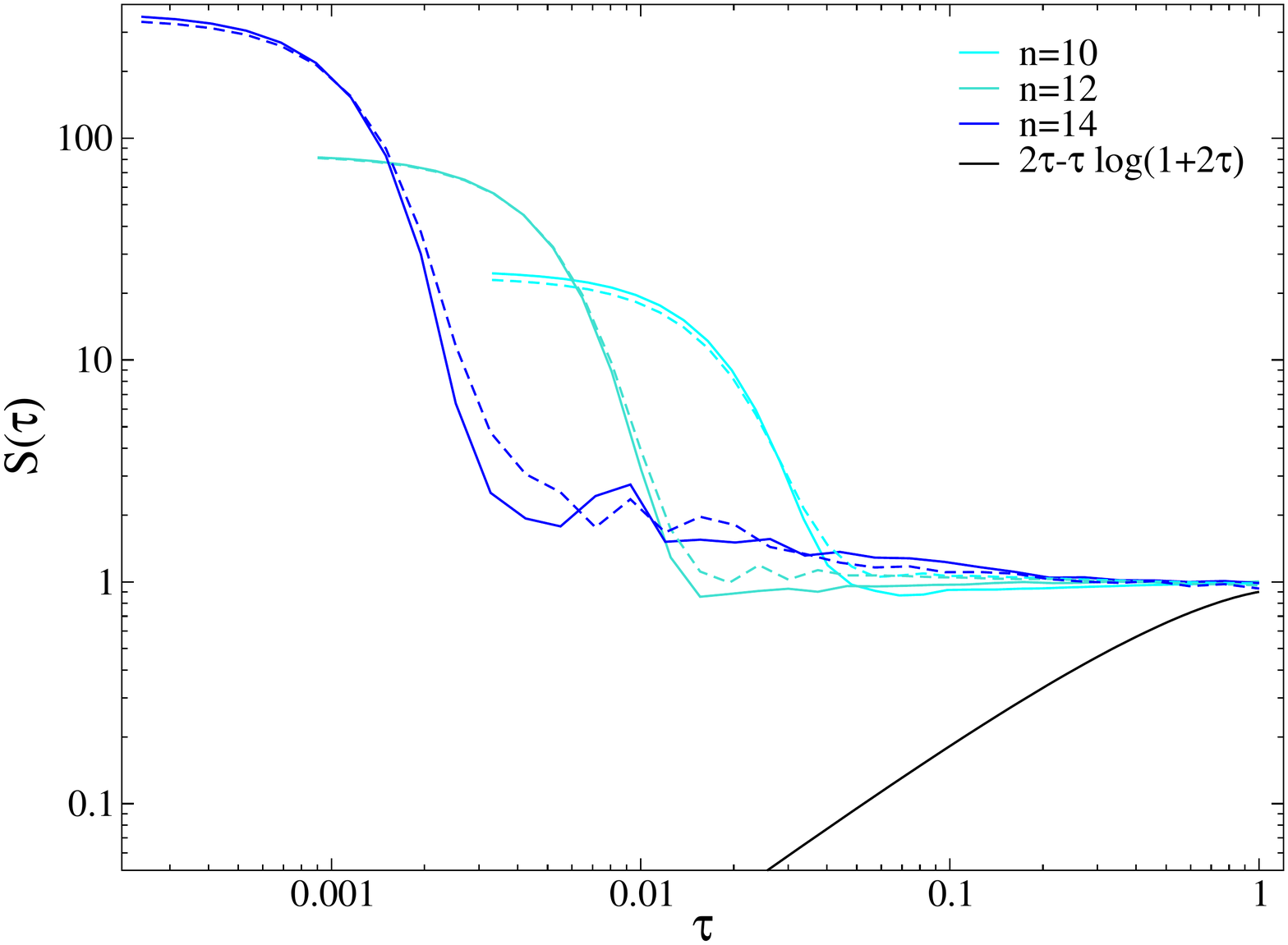}
\caption{\label{fig:sff}
SFF $S(\tau)$ for the Heisenberg XXZ random-field spin chain~(\ref{eq:Hxxz}) at $h=6$, deep into the MBL phase, and for $n=10$ (cyan), $n=12$ (turquoise), and $n=14$ (blue).
The dashed lines correspond to the SSF for the RP model~(\ref{eq:RP}), with $N = {{n}\choose{n/2}}$, $p(a) \propto e^{-E^2/(2 \sigma^2)}$, with $\sigma \propto \sqrt{n}$, $\gamma = 1.35$, and $\mu = 0.1$. The SFF of the RP model should approach the  GOE prediction, $S_{\rm GOE} (\tau) = 2 \tau - \tau \log(1 + 2 \tau)$ (black  line) in the thermodynamic limit.}
\end{figure}

{\bf (5)} This problem is in fact tightly related 
to the critical properties of the MBL transition.
It is well known that the FSA yields the mean-field exponent $\nu = 1$ for the divergence of the localization length at the Anderson transition irrespectively of the dimension and of the structure of the underlying graph~\cite{pietracaprina}.
The same exponent $\nu=1$ governs the transition from ergodic to multifractal eigenstates of the RP model taking place at $\gamma_c = 1$~\cite{pino}, while in the critical exponent for the generalized RP ensemble with log-normal distributed off-diagonal elements was recently found to vary between $1$ and $1/2$ depending on the parameters~\cite{kravtsov1}. However these critical behaviors are not compatible with the  
 the recent phenomenological RG studies for the MBL transition~\cite{KT,KT1} which instead predict a KT-like scenario with an exponential divergence of the localization length. 

{\bf (6)} A promising direction for future research would be to exploit the simplicity of our approach to address important questions such as 
the stability of MBL with respect to rare thermal inclusions of weak disorder that occur naturally inside an insulator and that may trigger a thermalization avalanche~\cite{avalanches}. 
It would be interesting for instance to insert by hand large but finite ergodic bubbles of weak disorder in the $1d$ XXZ spin chains and investigate the signature of quantum avalanches~\cite{avalanches,altman} by studying the effect of these bubbles on the scaling behavior of the amplitude of the propagators.

{\bf (7)} Finally, we would like to comment on the possible implications of our results on the recent debate on quantum chaos {\it vs} MBL~\cite{prosen,abanin}.
In fact a recent paper~\cite{prosen} has claimed that MBL is not a phase of matter, but rather a finite-size regime that yields to ergodic behavior in the thermodynamic limit. This conclusion was reached on the basis of a finite-size-scaling analysis of small $1d$ spin models using diagnostics from quantum chaos that probe the statistics of level spacing only, such as the structure form factor (SSF) $S(\tau)$ and the average level spacing ratio $\langle r \rangle$. 
In the light of the new interpretation proposed here, it is instructive to recall the known results for the statistics of eigenvalues of the RP model~\cite{kravtsov}:
In the intermediate regime $1 < \gamma <2$ of delocalized but nonergodic wavefunctions, although the average DoS asymptotically converges to the distribution of the  diagonal energies $p(a)$ (and not to the Wigner semicircle), the nearest-neighbor level statistics is described by the Wigner-Dyson statistics~\cite{kravtsov,facoetti}. In particular the (unfolded) spectral form factor was shown to be universal, i.e. independent of the specific form of $p(a)$, and to converge to the Wigner-Dyson form for $\gamma < 2$, and to Poisson only for $\gamma > 2$~\cite{kravtsov}. Similarly, $\langle r \rangle$ approach the GOE universal value $\langle r \rangle \approx 0.53$ in the thermodynamic limit for $1 < \gamma < 2$ and the Poisson one $\langle r \rangle \approx 0.38$ for $\gamma > 2$. In fact the $N^{D_1}$ states close in energy inside each mini-band exhibit level repulsion and GOE-like correlations. The crossover from GEO-like behavior to Poisson statistics occurs thus on the scale of the Thouless energy $\eta_{\rm th} = N^{D_1} \times \delta = N^{1-\gamma}$, which is vanishingly small for $N \to \infty$ but is still much larger than the mean level spacing.

As discussed in Sec.~\ref{sec:ldos}, the behavior of the spectral statistics of the many-body problem exhibits several important differences with respect to the RP model. The fact that the crossover scale $\eta_\star$ below which one observes  the characteristic localized behavior (see fig.~\ref{fig:ldos})  is much larger than the mean level spacing indicates that  consecutive energy levels are not hybridized by the off-diagonal perturbation, whose effect only sets in  on an larger energy scale $\gg \delta$, thereby implying that the level statistics on the scale of the mean level spacing should be of Poisson type.
Moreover, differently from the RP model, for the many-body Hamiltonian~(\ref{eq:Hxxz}) the mini-band in the LDoS are multifractal.
Yet,  if one takes the mapping to the RP model seriously beyond the qualitative level, one might be tempted to argue that any observable related to the level statistics on the scale of the mean level spacing only might be {\it uninformative} on the existence of a MBL transition in the thermodynamic limit. In order to illustrate this idea, we have measured the spectral form factor (SFF) defined as~\cite{prosen}:
\[
S(\tau) = \frac{1}{Z} \left \langle \left \vert \sum_{\beta = 1}^{\cal V} g(\tilde{E}_\beta) e^{- i \tilde{E}_\beta t} \right \vert^2 \right \rangle \, ,
\]
where $\tilde{E}_\beta$ are the unfolded eigenvalues of~(\ref{eq:Hxxz}), such that $\langle \tilde{E}_{\beta + 1} - \tilde{E}_\beta \rangle = 1$, $g(\tilde{E}_\beta)=\exp[- (\tilde{E}_\beta - \tilde{E}_0)^2/(2 \lambda^2 \sigma^2_{\tilde{E}})]$ is a Gaussian filter ($\lambda = 0.2$ is a dimensionless parameter that controls the effective fraction of eigenstates included in the $S(\tau)$, $\tilde{E}_0$ and $\sigma^2_{\tilde{E}}$ are the average energy and the variance of the unfolded eigenvalues, respectively, for a given disorder realization), and the normalization $Z = \langle \sum_\beta \vert g(\tilde{E}_\beta) \vert^2 \rangle$ is such that $S(\tau) \to 1$ for $\tau \to \infty$. The numerical results for $S(\tau)$ 
computed from exact diagonalizations of small ($n=10$, $12$, and $14$) XXZ chains~(\ref{eq:Hxxz})  and for $h=6$, deep into the MBL phase, are shown in Fig.~\ref{fig:sff}.
Inspired by the analogy with the RP random matrix ensemble discussed above, we benchmark these results with the SFF obtained for the RP model~(\ref{eq:RP}) with parameters chosen in such a way to mimic as closely as possible the interacting one: We set $N = {{n}\choose{n/2}}$ (equal to the size of the HS of the interacting model), and consider a Gaussian distribution of diagonal energies $p(a)$ of variance $\propto n$ and equal to the variance of the random energies $E_i$ in the spin configuration basis of the XXZ spin chain; We set $\gamma=1.35$ (which is approximately the value of the exponent found numerically at $h=6$, see Fig.~\ref{fig:gamma}), and $\mu = 0.1$ (which is the value of the prefactor in Eq.~(\ref{eq:Gtyp}) obtained by fitting the numerical data of Fig.~\ref{fig:FSA} at $h=6$).
The SSFs computed for the RP model (dashed lines in Fig.~\ref{fig:sff}) turn out to be very similar to the ones of the interacting model at the same value of $n$, although we know rigorously that they should approach the GOE result $2 \tau - \tau \log(1 + 2 \tau)$ (black line) for $N \to \infty$. However, as shown by Fig.~\ref{fig:sff}, the convergence to the GOE asymptotic result is very slow and finite-size effect are very big at finite $N$ due to the fact that the Thouless energy is still too close to the  mean level spacing.
All in all, this analysis suggests that one should take extreme caution when using diagnostics for the MBL transition based on the statistics of energy levels on the scale of the mean level spacing only, as they might slowly drift towards a GOE-like behavior in the thermodynamic limit even deep inside the MBL phase~\cite{abanin}.
This might also explain why the critical disorder estimated from the crossing of the curves of the average level spacing ratio $\langle r \rangle$ at different system sizes drifts to larger $h$ when increasing $n$~\cite{pal,prosen}.

\begin{acknowledgments}
	I would like to thank  G. Biroli, D. Facoetti, I. V. Gornyi,  I. Khaymovich, G. Lemari\'e, D. J. Luitz, A. D. Mirlin, M. Schir\'o, and V. Ros for many enlightening and helpful discussions.
\end{acknowledgments}

\appendix

\section{Results for the Imbrie model and for interacting fermions in a QP potential} \label{app}

In this appendix we provide more details and supplemental information related to several points discussed in the main text. In particular
we present the results obtained for two other models for the MBL transition described below.

\begin{figure}
	\includegraphics[width=0.468\textwidth]{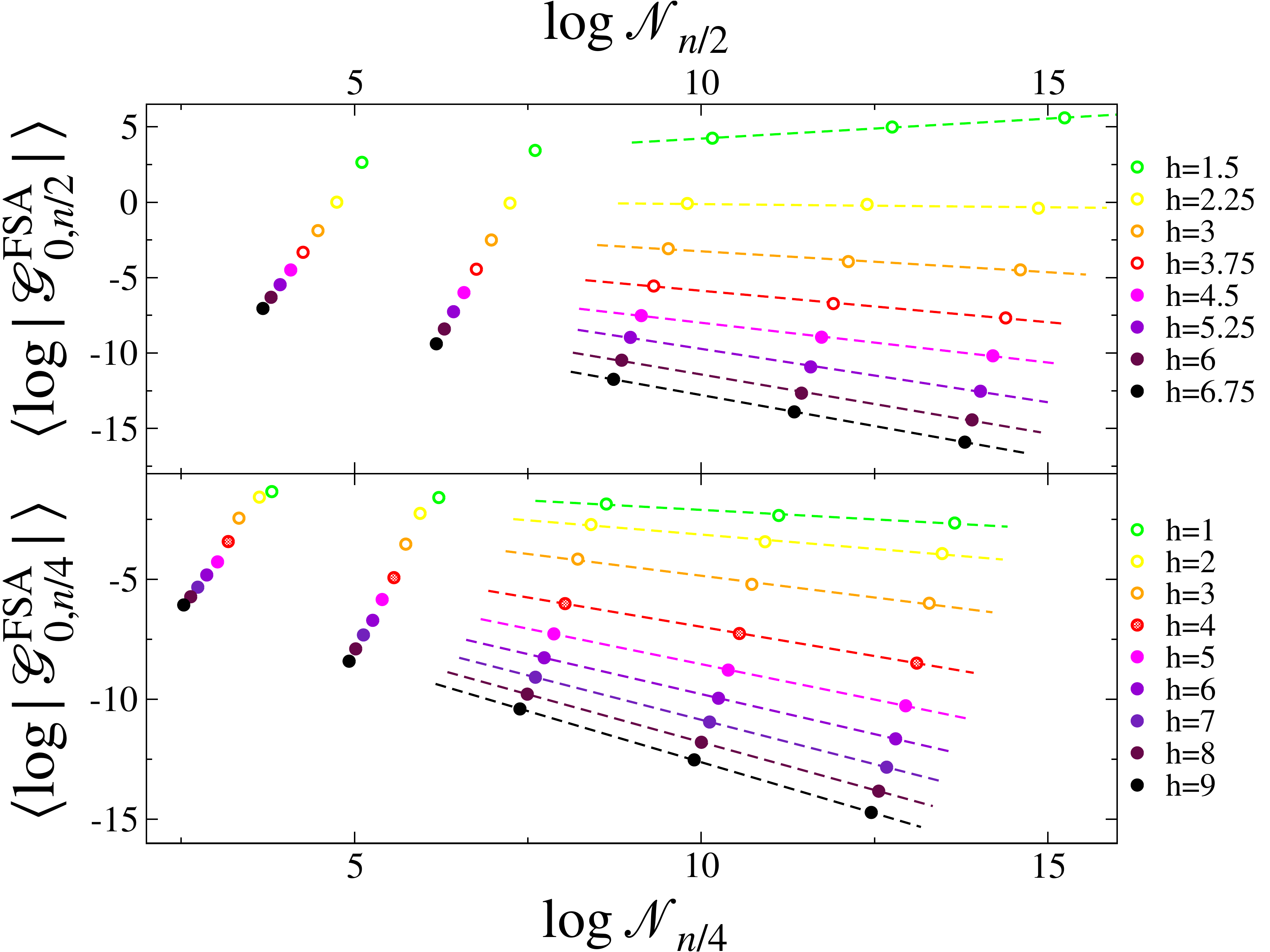}
	\caption{\label{fig:FSAIQP}
		Logarithm of the typical value of the propagator ${\cal G}_{q=0}^{\rm FSA}$ as a function of the log of the number of states ${\cal N}_{q=0}$ in the HS at zero overlap from $| 0 \rangle$ and in the same energy shell for the Imbrie model~(\ref{eq:HI})---top panel--- and for the QP model~(\ref{eq:Hqp})---bottom panel. $n$ varies from $12$ to $26$ for the Imbrie model and from $12$ to $28$ for the QP model. Different colors correspond to different values of the disorder across the MBL transition. The dashed lines correspond to linear fits of the numerical results of the form of Eq.~(\ref{eq:Gtyp}) at large $n$. Filled circles correspond to values of the disorder strength such that $\gamma>1$ ($h>h_c$), while empty circles to $\gamma <1$ ($h<h_c$).}
\end{figure}

\subsection{The models}

The  ``Imbrie'' model is defined by the following Hamiltonian:
\begin{equation} \label{eq:HI}
{\cal H}_I = \sum_{i=1}^n \left( \Delta_i \sigma_i^z \sigma_{i+1}^{z} + h_i \sigma_i^z + \Gamma_i \sigma_i^x \right) \, .
\end{equation}
We follow Ref.~\cite{abanin} and set $\Gamma_i=1$, and $\Delta_i$ and $h_i$ uniformly distributed in $\Delta_i \in [0.8,1.2]$ and $h_i \in [-h,h]$.
The existence of the MBL transition has been proven rigorously for this model under the minimal assumption of absence of level attraction~\cite{LIOMS}.
The numerical results of~\cite{abanin} seem to indicate that the critical disorder strength should be in the interval $h_c \in [3.75,4.5]$.

We also studied a one-dimensional model of spinless fermions on a QP lattice~\cite{barlev,huseQP,roscilde}:
\begin{equation} \label{eq:Hqp}
{\cal H}_{\rm QP} = \sum_{i=1}^n \left[ t \left( c_i^\dagger c_{i+1} + {\rm h.c.} \right) + \Delta n_i n_{i+1} + V_i n_i \right ] \, ,
\end{equation}
where $V_i$ is a QP potential of the form:
\[
V_i = h \cos (2 \pi \omega i + \phi) \, ,
\]
and $\phi$ is a random phase.
(Note that ${\cal H}_{\rm QP}$ exactly maps to the XXZ spin chain in a QP magnetic field under a Jordan-Wigner transformation.)
This model is very similar to the one realized in cold atom experiments~\cite{experiments1,experiments2,experiments3}.
As in Ref.~\cite{barlev}, we set $t=1$, the irrational number $\omega$ to be the inverse of the golden mean $\omega = (\sqrt{5} - 1)/2$, $\Delta = 1$, and only consider the half-filling sector $\sum_i n_i = n/2$. For this choice of the parameters previous studies~\cite{roscilde,barlev} have established the presence of a MBL transition with a critical strength of the QP potential in the interval $h_c \in [3.5,4.5]$.
For both models we consider periodic boundary conditions.

\subsection{Scaling of the matrix elements within the FSA}
In the following we discuss the results obtained for these two models by applying the same analysis described in Sec.~\ref{sec:FSA} of the main text. Concretely, we compute the probability distributions of the propagators
${\cal G}_{0,nx}^{\rm FSA}$ when the many-body systems are recast as single-particle tight binding problems~(\ref{eq:H1}). 

\begin{figure}
	\includegraphics[width=0.468\textwidth]{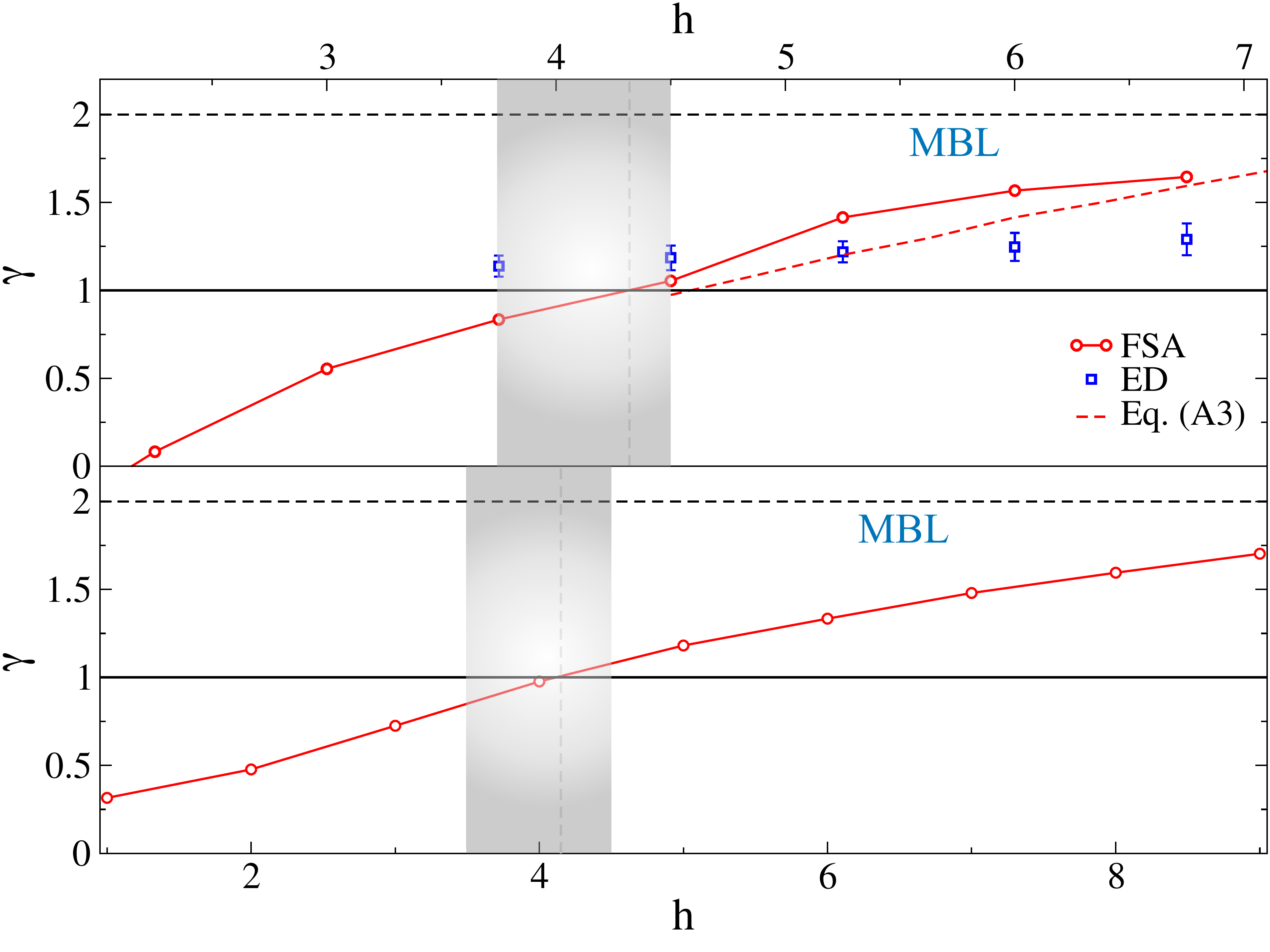}
	\caption{\label{fig:gammaIQP}
		Effective exponent $\gamma$ describing the scaling of typical value of the matrix elements ${\cal G}_{q=0}^{\rm FSA}$  with the number of configurations ${\cal N}_{q=0}$ at zero overlap with a given configuration in the HS, Eq.~(\ref{eq:Gtyp}), for the Imbrie model (\ref{eq:HI})---top---and for the QP model~(\ref{eq:Hqp})---bottom.
		Red circles correspond to the results obtained by linear fitting of $\langle \log | {\cal G}_{q=0}^{\rm FSA} | \rangle$ {\it vs} ${\cal N}_{q=0}$ at large $n$, 
		Fig.~\ref{fig:FSAIQP}.
		In the top panel the red dashed line corresponds to the strong disorder approximation for $\gamma$ given in Eq.~(\ref{eq:strong_disorderI}), while the blue squares give the estimation of $\gamma$ obtained by inspecting the unusual scaling of the LDoS computed via exact diagonalizations of~(\ref{eq:HI}), see Sec.~\ref{sec:ldosI}.
		The gray vertical shaded region marks the disorder range in which the MBL transition occurs according to the most recent numerical results (for chains of about the same size of the ones considered here)~\cite{abanin,barlev,huseQP,roscilde}, showing that  $\gamma_c = 1$ falls perfectly within it.
		}
\end{figure}

By choosing the spin configuration basis, the HS of~(\ref{eq:HI}) is a
$n$-dimensional hypercube of ${\cal V} = 2^n$ sites (the total magnetization is not conserved by ${\cal H}_I$).
Each configuration $| a \rangle = | \! \! \uparrow \downarrow \uparrow \! \cdots \rangle$ of $n$ spins corresponds to a corner  
of the  hypercube  by  considering $\sigma_i^z = \pm1$ as  the  top/bottom  face  of  the  cube's $n$-th  dimension. 
The random part of the Hamiltonian is by definition diagonal on this basis, and gives {\it correlated} 
random energies on each site orbital of the hypercube, $E_a = \langle a | \sum_{i=1}^n [ \Delta_i \sigma_i^z \sigma_i^{z+1} + h_i \sigma_i^z] | a \rangle$.
The interacting part of ${\cal H}_I$ acts as single spin flips on the configurations $\{ \sigma_i^z \}$, and plays the role the hopping rates connecting ``neighboring'' sites in the configuration space (with $t=1$).
At $\Gamma = 0$ the many-body eigenstates of (\ref{eq:HI}) are simply product states of the form $\vert \sigma_1^z \rangle \otimes \vert \sigma_2^z \rangle \otimes \cdots \otimes \vert \sigma_n^z \rangle$, and the system is fully localized in this basis.

Similarly, for the QP model we choose as a basis the tensor product of the simultaneous eigenstates of the number operators $n_i$, 
$| a \rangle = | \!\! \bullet \! \circ \! \bullet \! \cdots \rangle$, 
such that for $t=0$ the many-body eigenstates are perfectly localized. The HS of~(\ref{eq:Hqp}) is then represented by the same graph as for the XXZ random-field Heisenberg model considered in the main text, and its size is ${\cal V} = {{n}\choose{n/2}}$. The diagonal part of~(\ref{eq:Hqp})  yields the on-site quasi-random energies $E_a = \langle a | \sum_{i=1}^n [ \Delta n_i n_{i+1} + V_i n_i ] | a \rangle$, while the interacting part 
allows tunneling between ``neighboring'' configurations with hopping rate $t=1$.
The connectivity of the state $| a \rangle$ is equal to the number of pairs $\bullet \circ$ or $\circ \bullet$ 
in the state, and ranges from $2$ to $n$ with average value $\langle z \rangle \approx (n+1)/2$.

For both model we pick an infinite temperature many-body state $| 0 \rangle $  with energy $E_0$ in the middle of the spectrum, 
and determine the probability distribution of the matrix elements with configurations at distance $nx$ from it  at the lowest order in the hopping.
As for the XXZ model presented in the main text, instead of considering all $x$-sectors separately, for simplicity we only focus on the states at zero overlap 
from the initial one, i.e. when half of the spins have flipped or half of the particles have moved respectively. (Specifically the overlap is defined as $q = (1/n) \sum_i {}^0\sigma_i^z \cdot \sigma_i^z$ and $q = (1/n) \sum_i (n_i^0 - 1/2) (n_i - 1/2)$ for the two models, where the random initial state is denoted as $| 0 \rangle = \{ {}^0\sigma_i^z  \}$ and $| 0 \rangle = \{ n_i^0  \}$ respectively.) 
The shortest path to achieve $q=0$ corresponds to Hamming distance $n/2$ on the hypercube for the Imbrie model and to Hamming distance $n/4$ on the graph for the QP model.
The total number of configurations at $q=0$ from $| 0 \rangle$ and in the same energy shell is 
${\cal N}_{n/2} \approx {{n}\choose{n/2}} e^{S(0)} \delta E$ for the Imbrie model and ${\cal N}_{n/4} \approx {{n/2}\choose{n/4}}^2 e^{S(E_0)} \delta E$ for the QP model, where $S$ is the microcanonical entropy in the middle of the spectrum, defined as the logarithm of the number of states at that energy.

\begin{figure}
	\includegraphics[width=0.468\textwidth]{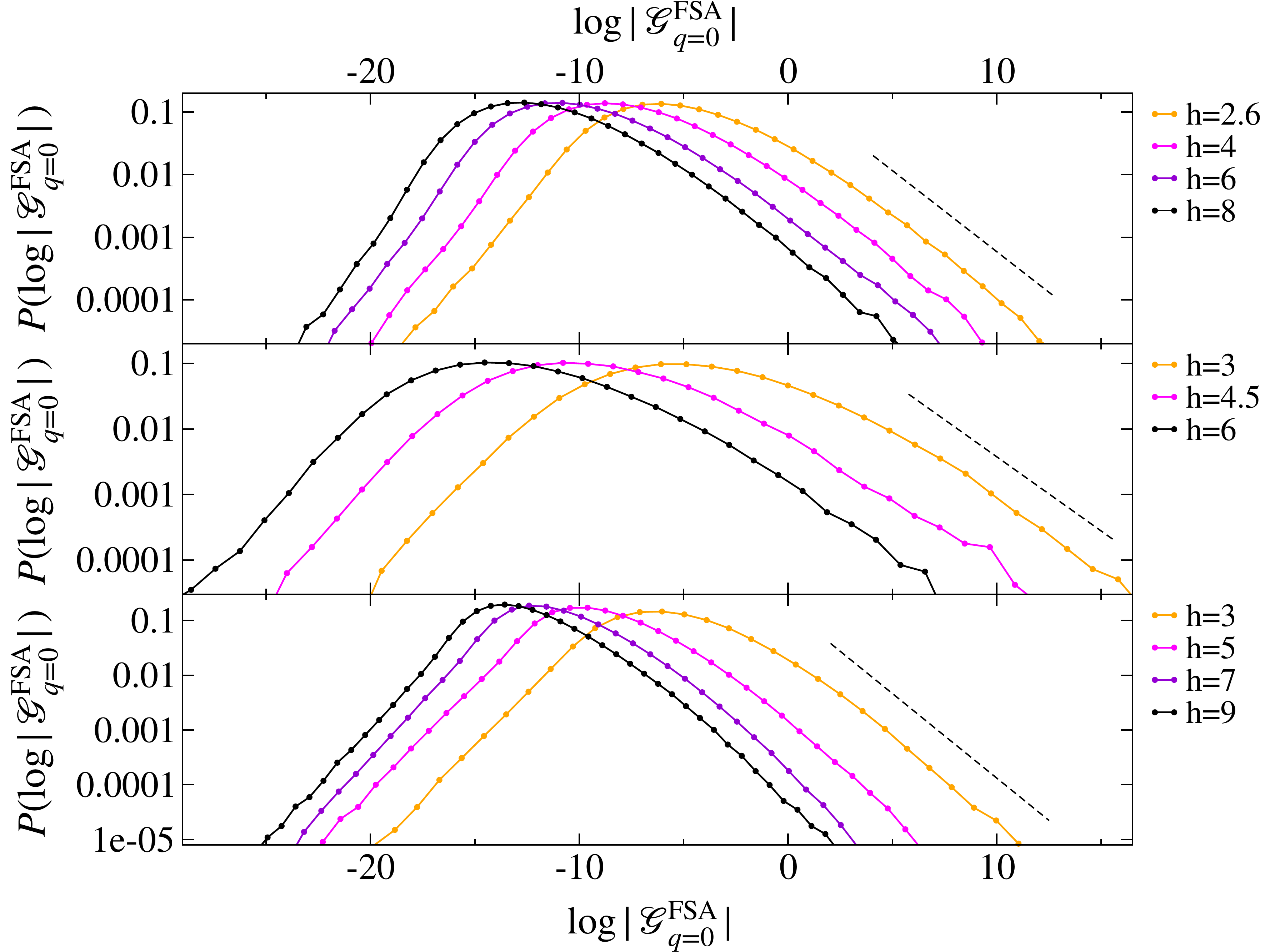}
	\caption{\label{fig:PG}
		Probability distribution of the logarithm of the amplitude of the propagator  (computed within the FSA) between states at extensive distance in the HS for the Heisenberg XXZ disordered spin chain (\ref{eq:Hxxz})---top---, for the Imbrie model (\ref{eq:HI})---middle---, and for the QP model (\ref{eq:Hqp})---bottom---, and for $n=24$. Several values of the disorder are shown across the MBL transition. The black dashed line give the power-law tails of the distributions, $P(|{\cal G}_{q=0}^{\rm FSA}|) \sim |{\cal G}_{q=0}^{\rm FSA}|^{-(1+\mu)}$, with $\mu \simeq 0.65$, $\mu \simeq 0.42$, and $\mu \simeq 0.61$ respectively.}
\end{figure}

In Fig.~\ref{fig:FSAIQP} we plot (the log of) the typical value of the propagator (computed within the FSA) as a function of (the log of) ${\cal N}_{q=0}$ for the two models varying the system size. Different curves correspond to different values of the disorder strength across the MBL transition~\cite{abanin,barlev,huseQP,roscilde}.
The plots clearly shows that Eq.~(\ref{eq:Gtyp}) holds. 
The value of the effective exponent $\gamma$ is obtained by the linear fitting of the numerical results at large ${\cal N}_{q=0}$ (dashed lines of Fig.~\ref{fig:FSAIQP}), and is plotted in
Fig.~\ref{fig:gammaIQP}. This figure shows that for both models
the critical disorder determined by the most recent numerical works is consistent with the value of the disorder such that $\gamma$ becomes larger than one, while $\gamma$ tends to $2$ in the limit of infinite disorder. This is exactly the same behavior found for the XXZ spin chain and discussed in the main text, Figs.~\ref{fig:FSA} and~\ref{fig:gamma}, supporting the robustness of our conclusions and the validity of the criterion  built on the FGR, $\Gamma_{\rm ergo} = {\cal N}_{q=0} |{\cal G}_{q=0}|_{\rm typ}^2 \to 0$, for the MBL transition.

Note that for the Imbrie model one can repeat the strong disorder approximation discussed in Sec.~\ref{sec:SD}, and straightforwardly obtain the following expression:
\begin{equation} \label{eq:strong_disorderI}
\gamma \underrel{n\to \infty}{\approx} \frac{\frac{1}{2}\log \frac{n}{2}! - \langle \log(2 h_1)\rangle  - \cdots - \langle \log(\sum_{i=1}^{n/2} 2 h_i)\rangle}{\log {{n}\choose{n/2}} - \frac{1}{2} \log n} \, , 
\end{equation}
which is in good agreement with the numerical results (red dashed curve in the top panel of Fig.~\ref{fig:gammaIQP}).
The Bethe lattice estimation of the ergodicity breaking transition based on the exponential decay of the matrix element on a single branch, Eq.~(\ref{eq:Wergo}), yield instead $h_c = 2 \sqrt{3} e \approx 9.4$.

\begin{figure}
	\includegraphics[width=0.54\textwidth]{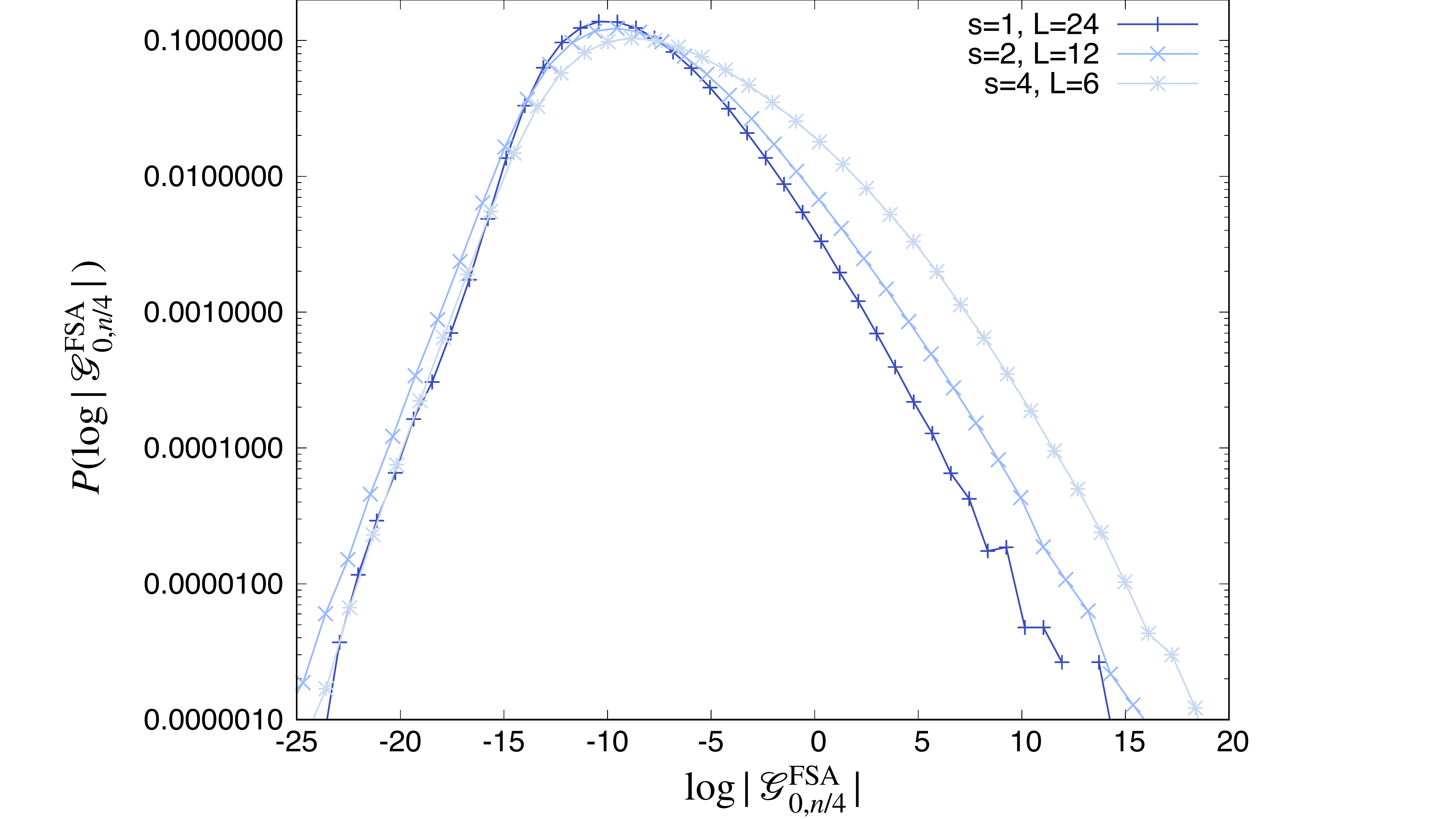}
	\caption{\label{fig:PG2d}
	Probability distribution of the logarithm of the amplitude of the propagator  (computed within the FSA) between states at extensive distance in the HS for the model of hard-core bosons with nearest neighbor interactions ~(\ref{eq:HCB}) for $h=6$ and total number of sites $n=Ls=24$, and for $s=1$, $2$, and $4$.}
\end{figure}

In Fig.~\ref{fig:PG} we show the full probability distributions of the amplitude of the tunneling rates for the three models considered in this work and for several value of the disorder strength across the MBL transition.
The distributions have power-law tails $P(|{\cal G}_{q=0}^{\rm FSA}|) \sim |{\cal G}_{q=0}^{\rm FSA}|^{-(1+\mu)}$ as expected for any value of $E$ in Eq.~(\ref{eq:FSA}) within the support of the probability distribution of the random energies $E_c$. The exponent $\mu$ varies from one model to another, but does not depend (or depends very weakly) on the disorder strength and on the system size.

Finally, in Fig.~\ref{fig:PG2d} we  plot the probability distribution of the propagator $P(|{\cal G}_{0,n/4}^{\rm FSA}|)$  for the quasi-$1d$ model of hard-core bosons with nearest neighbor interactions described by the Hamiltonian~(\ref{eq:HCB}), obtained when the width of the strip from $s=1$ to $s=4$. The curves correspond to disorder strength $h=6$, inside the MBL regime,  and for the same total number of sites $n=Ls =24$). One clearly observes a strong enhancement of the tails of the distribution when $s$ is increased, corresponding to rare large tunneling amplitudes, accompanied by a moderate increase of the typical value.

\begin{figure}
	\includegraphics[width=0.468\textwidth]{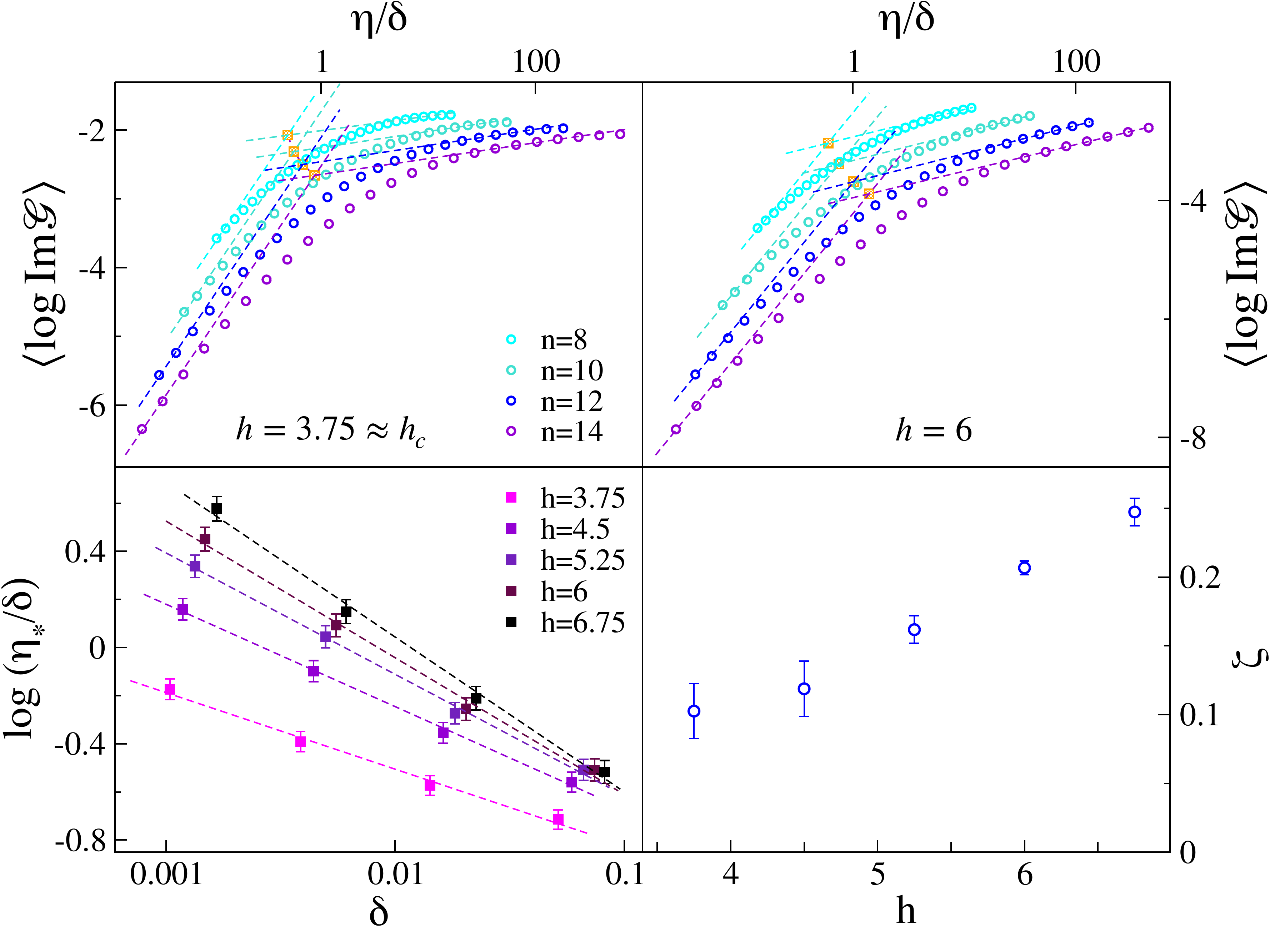}
	\caption{\label{fig:ldosI}
		Top panels: $\langle \log {\rm Im}{\cal G} (E_0 + i \eta) \rangle$ as a function of the imaginary regulator expressed in units of the mean level spacing, $\eta/\delta$, for the Imbrie model~(\ref{eq:HI}), for $h=3.75 \approx h_c$ (left) and $h=6$ (right), and for four chain length, $n=8$ (cyan), $n=10$ (turquoise), $n=12$ (blue), and $n=14$ (violet).
		The dashed straight lines represents the linear fits of $\langle \log {\rm Im}{\cal G} (E_0 + i \eta) \rangle$ as a function of $\log (\eta / \delta)$ at small and large $\eta$ with slope $1$ and $\zeta(h)$ respectively, Eq.~(\ref{eq:fitldos}). The orange squares mark the crossing points of the two straight lines which yield our estimation of the crossover scale $\eta_\star$. Bottom left:  Logarithm of the crossover scale $\eta_\star / \delta$ as a function of the mean level spacing $\delta$ for several values of the disorder strength $h \gtrsim h_c$ and four system sizes $n$ from $8$ to $14$. The dashed line correspond to fits of the data of the form $\eta_\star / \delta \propto \delta^{D_1 - 1}$, which gives an estimation of the effective exponent $\gamma = 2 - D_1$~\cite{kravtsov} (blue squares of Fig.~\ref{fig:gammaIQP}). Bottom right: Exponent $\zeta(h)$ describing the behavior of the typical value of the LDoS for $\eta \gg \eta_\star$ as a function of the disorder strength.
		}
\end{figure}

\subsection{Signature of  eigenstates' multifractality  in the spectral statistics of the Imbrie model} \label{sec:ldosI}
As done in Sec.~\ref{sec:ldos} for the XXZ random-field spin chain, one can investigate the signatures of the multifractality of the eigestates from the unusual scaling limit of the spectral statistics of the Imbrie model.
In the top panels of Fig.~\ref{fig:ldosI} we plot the logarithm of the typical value of the LDoS, $\langle \log {\rm Im}{\cal G} \rangle$, as a function of the imaginary regulator measured in units of the mean level spacing, $\eta/\delta$, for four system sizes ($n$ from $8$ to $14$) and for two values of the disorder strength $h$ across the MBL transition. These plots are obtained by inverting exactly the many-body Hamiltonian~(\ref{eq:HI}) in presence of the imaginary regulator, and averaging over several (about $2^{36-2n}$) independent realizations of the disorder. The curves show the existence of the crossover scale $\eta_\star$ separating the behavior of ${\rm Im}{\cal G}$ at small and large $\eta$, as described by Eq.~(\ref{eq:fitldos}). We have extracted such crossover scale from the data by applying the procedure described in Sec.~\ref{sec:ldos} of the main text. In the bottom-left panel of Fig.~\ref{fig:ldosI}
 we plot $\log(\eta_\star / \delta)$  as a function of $\delta$ for several values of $h \gtrsim h_c$.
We observe that $\log (\eta_\star / \delta)$ increases linearly by decreasing $\log \delta$ (i.e., increasing $n$), consistently with the presence of multifractal eigenfunctions which only  occupy a subextensive part of the HS.
By fitting $\eta_\star / \delta \propto \delta^{D_1 - 1}$ one obtains an independent estimation of the fractal dimension $D_1$ and, by analogy with the RP model, of the effective exponent $\gamma = 2 - D_1$~\cite{kravtsov} (blue squares in Fig.~\ref{fig:gammaIQP}).
For the Imbrie model, however, such numerical estimation of $\gamma$ 
does not agree well with the numerical results of the FSA. This discrepancy is possibly due to strong finite-size effects, since simple intuitive arguments suggest that one should find $\gamma \sim 2 - c/h$ at strong disorder.


\begin{thebibliography}{99}
\bibitem{RRGmulti} Note however that the Anderson model on the RRG is known to display a very peculiar form of strong multifractaliy in the whole localized phase~\cite{evers} for the wavefunction moments $\Upsilon_q$ with $q<q_\star = 0.5$, while $\Upsilon_q = 0$ for $q>q_\star$. This corresponds to the multiftactal behavior of very the small amplitudes of the exponentially localized eigenstates.
\bibitem{entanglement_D2} This behavior is closely related to the volume-law scaling of the long-time saturation value of the entanglement entropy for the case when the initial state is a basis state~\cite{LIOMS1,entanglement}.
\bibitem{caveat} Note that in fact the accessible volume  at a given extensive energy $E_0$ is ${\cal V} \Omega(E_0)$, where $\Omega(E_0) = e^{S(E_0)}$ is the many-body density of states, given by the exponential of the microcanonical entropy, $S(E_0)$.
\bibitem{disclaimer} Note that here we are implicitly generalizing the results obtained at $q=0$ (i.e., Hamming distance $n/4$ on the HS for the XXZ model) to all other $x$-sector, assuming that the same effective exponent $\gamma$ describes the scaling of the matrix elements at all distances, while one should instead compute the exponent $\gamma$ in each sector separately, and verify that ergodicity is broken for all $x$.
\bibitem{disclaimer1} Note that we are implicitly assuming that the loops of the Bethe lattice (if any) are much larger than $n$. This means that the total number of sites must be such that $N \gg k^n$.
\end{thebibliography}
\end{document}